\documentclass[12pt,preprint]{aastex}
\usepackage{enumerate}
\usepackage{float}
\usepackage{placeins}
\usepackage{amsmath}
\usepackage{natbib}
\usepackage{color}
\usepackage[utf8x]{inputenx}
\usepackage[greek,english]{babel}
\usepackage{footnote}
\usepackage{pdflscape}

\newcommand{\HII}{H\,{\sc ii}~}

\defcitealias{Nicholls12}{NDS12}
\defcitealias{Palay12}{PNPE12}

\begin{document}

\title{Measuring nebular temperatures: the effect of new collision strengths with equilibrium and $\kappa$--distributed electron energies}
\shorttitle{The effect of non-equilibrium electron energies}
\shortauthors{Nicholls et~al.}

\author{ David C. Nicholls\altaffilmark{1}, Michael A. Dopita\altaffilmark{1}\altaffilmark{2}, Ralph S. Sutherland\altaffilmark{1}, Lisa J. Kewley\altaffilmark{1} \& Ethan Palay\altaffilmark{3}}
\email{david@mso.anu.edu.au}
\altaffiltext{1}{Research School of Astronomy and Astrophysics, Australian National University, Cotter Rd., Weston ACT 2611, Australia }
\altaffiltext{2}{Astronomy Department, King Abdulaziz University, P.O. Box 80203, Jeddah, Saudi Arabia}
\altaffiltext{3}{Department of Astronomy, The Ohio State University, Columbus, OH 43210, USA}
\begin{abstract}
In this paper we develop tools for observers to use when analysing nebular spectra for temperatures and metallicities, with two goals: to present a new, simple method to calculate equilibrium electron temperatures for collisionally excited line flux ratios, using the latest atomic data; and to adapt current methods to include the effects of possible non-equilibrium ``$\kappa$'' electron energy distributions.  Adopting recent collision strength data for [O~{\sc iii}],   [S~{\sc iii}],  [O~{\sc ii}],  [S~{\sc ii}], and  [N~{\sc ii}], we find that existing methods based on older atomic data seriously overestimate the electron temperatures, even when considering purely Maxwellian statistics.   If $\kappa$ distributions exist in \HII regions and planetary nebulae as they do in solar system plasmas, it is important to investigate the observational consequences. This paper continues our previous work on the $\kappa$ distribution \citep{Nicholls12}. We present simple formulaic methods that allow observers to (a) measure equilibrium electron temperatures and atomic abundances using the latest atomic data, and (b) to apply simple corrections to existing equilibrium analysis techniques to allow for possible non-equilibrium effects. These tools should lead to better consistency in temperature and abundance measurements, and a clearer understanding of the physics of \HII regions and planetary nebulae.
\end{abstract}

\section{Introduction}

Fundamental to all methods of measuring temperatures and abundances in gaseous nebulae are the atomic data for the ionized nebular species.  In particular, an accurate knowledge of the collision strengths for the excitation of ionized nebular species is critical to obtaining reliable information on the conditions in these plasmas.  Unfortunately, computing these collision strengths is a lengthy and complex process, placing considerable demands on computational power.  Many current nebular abundance analysis methods make use of atomic data computed over 20 years ago.  In this work we assemble the best available modern data to investigate the effects on temperature and abundance measurement.  We find that the latest data makes a considerable difference to the answers obtained. 

All previous approaches have used ``effective collision strengths'', where the detailed computed collision strengths are convolved with Maxwell-Boltzmann electron energy distributions at fixed temperatures. In this work, we use the detailed collision strengths whose energy dependence has not been convolved with an electron energy distribution. Our approach has enabled us to build simple formulas which will allow the observer to calculate (equilibrium)  electron temperatures, based on the most recent atomic data.

We also return to the subject of our previous paper  \citep[hereafter, NDS12]{Nicholls12}, the non-equilibrium $\kappa$ electron energy distribution.   These distributions have been widely detected in solar system plasmas \citep{Pierrard10}, and \cite{Tsallis95} have explained from entropy considerations why and how such distributions can occur. As previous analyses have assumed equilibrium energy distributions in \HII regions and planetary nebulae, we revisit our reasons for considering non-equilibrium electron energy distributions in these objects.

In this paper we take the exploration of the $\kappa$ distribution further.  Using the un-convolved collision strengths, we explore in detail the effects of the $\kappa$ distribution. We derive formulae to simplify calculating the effects of a $\kappa$ distribution from conventional equilibrium results. In this way, observers can investigate the effect of any $\kappa$-type divergence from equilibrium electron energies.  

Our aim is to provide the observer with a set of tools to (a) take advantage of the latest atomic data for equilibrium calculations; and (b) using the $\kappa$ electron energy distribution, to correct apparent temperatures measured from temperature sensitive line ratios or recombination continua for subsequent abundance analyses. 

The paper is organised as follows. In section 2, we present a rationale for considering non-equilibrium electron energy distributions in gaseous nebulae. In section 3, we describe the $\kappa$ distribution for electron energies and compare the collisional excitation rates for the $\kappa$ and Maxwell-Boltzmann (M-B) distribution. In section 4, we discuss the factors involved in obtaining accurate collisionally excited line (CEL) equilibrium electron temperatures from theoretical collision strengths. In particular we point out the errors resulting from inaccuracies in the collision strengths used as the bases for most current direct electron temperature techniques.  We show that for non-equilibrium electron energy distributions, it is necessary to use detailed collision strengths for atomic species of interest, as distinct from the thermally averaged  \em effective\em\ collision strengths that are usually published; we discuss the effect of premature truncation of collision strength computation at high energies; and list the sources for the collision strength data we have used.  In section 5 we explore the effect of the $\kappa$ distribution on recombination processes, and how to calculate the effects of a $\kappa$ distribution on the apparent temperature and density of the recombining electrons; and we calculate the degree to which recombination lines are enhanced by the $\kappa$ distribution. In section 6, we describe in detail the effect of the $\kappa$ distribution on collisionally excited lines. Using a general expression for the collisional excitation rate ratio between the $\kappa$ and M-B distributions, we derive the relative intensity enhancements for different atomic species, and detailed equations for temperature sensitive line flux ratios. We show typical flux ratio vs. kinetic temperature plots for [O~{\sc iii}] and [S~{\sc iii}] for a range of values of $\kappa$, based on direct calculation from recently published collision strengths. 

Section 7 is the main focus of the paper, where we present a new, simple method for calculating equilibrium electron temperatures from line flux ratios using the most recent collision strength data, including density corrections; tools for measuring true (kinetic) CEL electron temperatures, using conventionally calculated equilibrium electron temperatures as a starting point; and a simple linear equation for converting between conventional measurement results and $\kappa$-corrected temperatures. In section 8 we discuss briefly the effect of $\kappa$ on strong line methods.  In section 9 we present ways to determine $\kappa$ and point out the need for and progress with implementing $\kappa$ effects in photoionization modelling codes.  In section 10 we summarize our conclusions. In Appendix 1 we list the temperature-sensitive lines for the most common atomic species found in \HII regions and PNe; the transition probabilities for these transitions; and the various factors appearing in the formulae in Section 5 which allow the temperature-sensitive line ratios to be computed for any internal energy temperature and value of $\kappa$. 

\section{Rationale for considering non-equilibrium electron energies.}

It has long been held that the electrons in \HII regions and planetary nebulae are in thermal equilibrium.  Analytical calculations of electron velocity distributions in gaseous nebulae were presented by \cite{Bohm47}.  Their work led them to state that the velocity distribution is ``very close to Maxwellian''.  \citet[][Ch.5]{Spitzer62}  also examined the thermalization process for electron energies in plasmas and found that electron energies equilibrate rapidly through collisions. This early work has lead later authors to assume the electrons in gaseous nebulae are always in thermal equilibrium.  However, Spitzer's analysis showed that the equilibration time of an energetic electron is proportional to the cube of the velocity, so even using equilibrium theory, plasmas with very high energy electrons take much longer to equilibrate than those excited by normal UV photons from stars found in \HII regions.

In more recent times, the electron energies in solar system plasmas have been measured directly by satellites and space probes.  This began with \cite{Vasyliunas68}, who found that the electron energies in the Earth's magnetosphere departed substantially from the Maxwellian, and resembled a Maxwellian with a high energy power law tail.  He showed that this distribution could be well described by what he called the ``$\kappa$ distribution''.  Since then, $\kappa$ distributions have been widely detected in solar system plasmas and are the subject of considerable interest in solar system physics.\footnote{Over 400 papers on the applications of $\kappa$  distributions in astrophysics had been published prior to 2009 \citep{Livadiotis09} and over 5000 in physics in general had been published prior to 2011 \citep{Livadiotis11b}.} They have been detected in the outer heliosphere, the magnetospheres of all the gas-giant planets, Mercury, the moons Titan and Io, the Earth's magnetosphere, plasma sheet and magnetosheath and the solar wind (see references in \cite{Pierrard10}). There is also evidence from IBEX observations that energetic neutral atoms in the interstellar medium, where it interacts with the heliosheath, exhibit $\kappa$ energy distributions \citep{Livadiotis11a}.  In solar system plasmas, the $\kappa$  distribution is the norm, and the MB distribution is a rarity. So we are confronted with the fact that despite the early theoretical work suggesting that the electrons in such plasmas should be in thermal equilibrium, they are almost always not.

Initially, $\kappa$ distributions were used as empirical fits to observed energies, and were criticized as  lacking a theoretical basis. Subsequently, the distribution has been shown to arise naturally from entropy considerations. See, for example, \citet{Tsallis95, Treumann99, Leubner02}, and the comprehensive analysis by \citet{Livadiotis09}.  They have explored ``q non-extensive statistical mechanics'' and have shown that $\kappa$ energy distributions arise as a consequence of this entropy formalism, in the same way as the Maxwell-Boltzmann distribution arises from Boltzmann-Gibbs statistics.  The requirement for this to occur is that there be macroscopic interactions between particles, in addition to the shorter-range Coulombic forces that give rise to Maxwell-Boltzmann equilibration.  Tsallis statistics provide a sound basis for the overtly successful use of the $\kappa$ distribution in describing solar system plasmas.  $\kappa$  distributions appear to arise whenever the plasma is being pumped rapidly with high energy non-thermal electrons, so that the system cannot relax to a classical Maxwell-Boltzmann distribution. \cite{Collier93} has also shown that $\kappa$-like energy distributions can arise as a consequence of normal power-law variations of physical parameters such as density, temperature, and electric and magnetic fields.

It is plausible that such conditions are also present in \HII regions and PNe---solar system plasma parameters span the many of the conditions found in gaseous nebulae, and, as in the solar system, \HII region plasmas can be magnetically dominated \citep{Arthur11, Nicholls12}---so it is important to investigate the effects of non-equilibrium energy distributions with high-energy  tails in occurring in gaseous nebulae, should they occur. 

Such non-Maxwellian energies may occur whenever the population of energetic electrons is being pumped in a timescale shorter than, or of the same order as the normal energy re-distribution timescale of the electron population. Suitable mechanisms include magnetic reconnection followed by the migration of high-energy electrons along field lines, the development of inertial Alfv\'en waves, local shocks (driven either by the collision of bulk flows or by supersonic turbulence), and, most simply, by the injection of high-energy electrons through the photoionization process itself. Normal photoionization produces supra-thermal electrons on a timescale similar to the recombination timescale. However, energetic electrons can be generated by the photoionization of dust \citep{Dopita00}, and X-ray ionization can produce highly energetic ($\sim$ keV) inner-shell (Auger process) electrons (e.g. \citet{Shull85, Aldrovandi85, Petrini97}, and references therein). These photoionization-based processes should become more effective where the source of the ionizing photons has a ``hard'' photon spectrum. Thus, the likelihood of the ionized plasma having a $\kappa$ electron energy distribution would be high in the case of either photoionization by an Active Galactic Nucleus (AGN), or the case of PNe, where the effective temperature of the exciting star could range up to $\sim 250,000$K.

So we have no shortage of possible energy injection mechanisms capable of feeding the energetic population on a timescale which is short compared with the collisional re-distribution timescale. The rate of equilibration falls rapidly with increasing energy, and we would expect there to be a threshold energy above which any non-thermal electrons have a long residence time. These can then feed continually down towards lower energies through conventional collisional energy redistribution, thus maintaining a  $\kappa$ electron energy distribution. 

In addition to the energy injection mechanisms capable of maintaining the excitation of  suprathermal distributions, several authors (\citet{Livadiotis11b} and references therein; \citet{Shizgal07, Treumann01}) have investigated the possibility that the $\kappa$ distribution may remain stable against equilibration longer than conventional thermalization considerations would suggest.  In particular, distributions with $2.5 \gtrsim \kappa > 1.5$---detected, for example, in Jupiter's magnetosphere---appear to have the capacity, through increasing entropy, of moving to values of lower $\kappa$ \citep{Livadiotis11b} i.e. away from (Maxwell-Boltzmann) equilibrium.  While the physical application of this aspect of $\kappa$ distributions remains to be explored fully, it suggests that where q non-extensive entropy conditions operate, the suprathermal energy distributions produced exist in ``stationary states'' where the behaviour is, at least in the short term, time-invariant \citep{Livadiotis10a}.  These states may have longer lifetimes than expected classically.  This is consistent with the numerous observations in solar system plasmas, that $\kappa$ electron and proton energy distributions are the norm. 

It is likely, therefore, that photoionized plasmas in gaseous nebulae will show departures from a Maxwell distribution to some degree. The key questions are, is this important, and does it produce observable effects in the nebular diagnostics which we have relied upon hitherto?

The answer to both questions appears to be `yes'. For several decades, systematic discrepancies have plagued abundance measurements derived from observations of emission lines and emission continua in \HII regions and PNe.  In particular, abundances determined from collisionally excited lines (CEL) for different ions differ from one another, and temperatures determined from Hydrogen and Helium bound-free continuum spectra are consistently lower than those obtained from CELs. As a consequence, chemical abundances determined from the optical recombination lines (ORL) are systematically higher than those determined from CELs.  These discrepancies are often referred to as the ``abundance discrepancy problem'' and are sometimes even parameterized as the ``abundance discrepancy factor'' (ADF). The problem was first observed 70 years ago and has been discussed regularly in the literature for 40 years.  See, for example, \citet{Wyse42, Peimbert67, Liu00, Stasinska04, Garcia07}.

A number of attempts have been made to explain these differences.  The earliest attempt appears to be by \citet{Peimbert67}, who proposed small temperature inhomogeneities through the emitting regions as the cause.  Later, \citet{Liu00} suggested the presence of a two-phase ``bi-abundance'' structure, where the emitting regions contain cool, metal-rich, hydrogen poor inclusions. However, neither explanation appears to be fully satisfactory: the temperature fluctuation model often requires large fluctuations to explain the observed discrepancies, without suggesting how these fluctuations could arise. The bi-abundance model requires proposing inhomogeneities where, in some cases, none are observed, or where the physical processes militate against the stability of such inhomogeneities.  The reader is referred to the detailed discussion by \citet{Stasinska04}. Further, in neither of these mechanisms is the discrepancy between different CEL species explained. More recently, \citet{Binette12} have suggested that shock waves may contribute to the apparent discrepancies, but they state that the mechanism needs to be explored further before it can be considered an explanation.  A common feature of all these approaches is that they assume the electrons involved in collisional excitation and recombination processes are in thermal equilibrium. 

In our previous paper \citepalias{Nicholls12} we showed that a non-equilibrium $\kappa$ electron energy distribution is capable of explaining both the ORL/CEL discrepancy, and the differences between electron temperatures obtained using different CEL species. The mechanism has been shown, for example, to provide an explanation in the case of [O~{\sc iii}] and [S~{\sc iii}] CEL lines \citep{Binette12}.  It is interesting to note that extreme departures from an equilibrium electron energy distribution are not required to accomplish this, and if there is pumping of electron energies by mechanisms clearly likely to occur in gaseous nebulae, such distributions may not be difficult to achieve.

In this paper, we continue to explore the implications of $\kappa$ energy distributions, using recently published collision strength data for key nebular species to model the effects the $\kappa$ distribution will have, if present, on the physics of \HII regions and PNe.

\section{The $\kappa$ distribution}

The $\kappa$ distribution resembles the M-B distribution at lower energies but has a high energy power law tail. Expressed in energy terms, the $\kappa$ distribution is \citepalias{Nicholls12}:
\begin{equation}\label{e1}
n(E) dE=\frac{2 N_e}{\sqrt\pi} \left( \frac{\Gamma(\kappa+1)}{(\kappa-\frac{3}{2})^{3/2} \Gamma(\kappa-\frac{1}{2})} \right) \frac{\sqrt E}{(k_BT_U)^{3/2} (1 + E/ \left[(\kappa-\frac{3}{2}) k_BT_U \right])^{\kappa + 1}} dE\ .
\end{equation}

The parameter $\kappa$ describes the extent to which the energy distribution differs from the M-B.  Its values lie in the range [$\frac{3}{2}, \infty$]. In the limit as $\kappa\to\infty$, the energy distribution reduces to the equilibrium M-B distribution:
\begin{equation}\label{e2}
n(E) dE=\frac{2 N_e}{\sqrt\pi} \frac{\sqrt E \ \exp\left[-E/k_BT_U\right]}{(k_BT_U)^{3/2}} dE\ .
\end{equation}
where $T_U$ is the ``kinetic'' or ``internal energy'' temperature, defined in terms of the energy density of the system, as per \citetalias[equation 5]{Nicholls12}; $N_e$ is the electron density; and $k_B$ is the Boltzmann constant. For a M-B energy distribution, $T_U$ is simply the thermodynamic temperature. Thus the Maxwell-Boltzmann distribution is a special case of the $\kappa$ distribution, where there is no long-range pumping of electron energies at timescales similar to the collisional relaxation time.

It can readily be shown by integration with respect to energy between the limits [0,$ \infty$]  that the area under the curves given in equations \eqref{e1} and \eqref{e2} is $N_e$, the electron density, in both cases, and in the case of $\kappa\to\infty$ the internal energy temperature is identically equal to the classical electron temperature. 

As shown by \citetalias{Nicholls12}, the collisional excitation rate from level 1 to level 2 for an M-B distribution is given by
\begin{equation}\label{e3}
R_{12}(\mathrm{M-B})=n_e N_e \frac{h^2 }{4 \pi^{3/2} m_e g_1} \left(k_B T_U \right)^{-3/2} \int\limits_{E_{12}}^\infty \Omega_{12}(E)\ \mathrm{ exp} \left[-\frac{E}{k_BT_U}\right] dE\ ,
\end{equation}
and for a $\kappa$-distribution, the corresponding rate is:
\begin{equation}\label{e4}
R_{12}(\kappa)=n_e N_e \frac{h^2}{4 \pi^{3/2} m_e g_1} \frac{\Gamma(\kappa+1)}{(\kappa-\frac{3}{2})^{3/2}\Gamma(\kappa-\frac{1}{2})} \left({k_BT_U}\right)^{-3/2}  \int\limits_{E_{12}}^\infty \frac{\Omega_{12}(E)}  {(1 + E/[(\kappa-\frac{3}{2}) k_BT_U])^{\kappa + 1}} \mathrm{d}E\ .
\end{equation}
where $\Omega_{12}$ is the collision strength for collisional excitations from level 1 to level 2, $E_{12}$ is the energy gap between levels 1 and 2, $g_1$ is the statistical weight of the lower state, and $\Gamma$ is the gamma function.

As a first order approximation, we can assume that the collision strength from excitations from level 1 to 2, $\Omega_{12}$, is independent of energy. For this case the ratio of the rates of collisional excitation from level 1 to level 2 for a $\kappa$ distribution can be expressed analytically \citepalias{Nicholls12} as:
\begin{equation}\label{e5}
\dfrac{R_{12}(\kappa)}{R_{12}(\mathrm{M-B})}=  \frac{\Gamma(\kappa+1)}{(\kappa-\frac{3}{2})^{3/2}\Gamma(\kappa-\frac{1}{2})}\left(1-\frac{3}{2\kappa} \right)\exp \left[\frac{E_{12}}{k_BT_U}\right] \left( 1+\dfrac{E_{12}}{(\kappa-\frac{3}{2})k_BT_U} \right) ^{- \kappa}\ .
\end{equation}
Detailed plots and values for this equation for a range of values of $\kappa$ are given in \citetalias[][Figure 5 and Table 1]{Nicholls12}.

Electron temperatures are generally measured using the line ratio of two emission lines with well-separated excitation energies, of which the best known is the $\lambda\lambda 4363/5007$ ratio for [O~{\sc iii}]. As shown in \citetalias[equations 12 and 13]{Nicholls12}\footnote{Note that there was an error in \citetalias[equation 12]{Nicholls12}, with a factor of $\sqrt{2/m_e}$ missing.  This omission disappears in the ratio process, however.}, for a M-B electron energy distribution, considering a simplified three-level atom, the ratio of the collisional excitation rate from level 1 to level 3 to the rate from level 1 to level 2, for the constant $\Omega$ case,  is given by the well-known formula:
\begin{equation}\label{e6}
\frac{R_{13}}{R_{12}}=\frac{\Omega_{13}}{\Omega_{12}}\ \exp\left[-\frac{E_{23}}{k_BT_U}\right]\ .
\end{equation}
where the collision strengths are once again considered to be independent of energy.

For a $\kappa$ electron energy distribution, again for the constant $\Omega$ case, the collisional excitation rate ratio is given by:
\begin{equation}\label{e7}
\frac{R_{13}}{R_{12}}=\frac{\Omega_{13}}{\Omega_{12}}\ \left[\frac{E_{13} + (\kappa-\frac{3}{2})k_BT_U}{E_{12} + (\kappa-\frac{3}{2})k_BT_U}\right]^{-\kappa}\ ,
\end{equation}
where $T_U$ is the kinetic or internal energy temperature.

\section{Collision strength considerations}
\subsection{``Non-averaged'' and effective collision strengths}

Equations \eqref{e3} and \eqref{e4} emphasise the importance of a knowledge of the collision strength over all energies.  In all the current literature, a M-B distribution has been assumed, and the effective collision strengths used are the collision strengths averaged over M-B energy distributions at different temperatures.  It should be noted that this averaging process is calculated for a fixed population of electrons, $N_e$.  Thus the full equation for deriving the effective collision strengths, $\Upsilon_{12}$, from the collision strengths, $\Omega_{12}$, for collisional excitations from level 1 to level 2 is:
\begin{equation}\label{e8}
\Upsilon_{12}(T) = \dfrac{\int\limits_{E=E_{12}}^\infty \Omega_{12}(E)  exp\left(\frac{-E}{kT}\right)d\left(\frac{E}{kT}\right)}{\int\limits_{E=E_{12}}^\infty exp\left(\frac{-E}{kT}\right)d\left(\frac{E}{kT}\right)}\ .
\end{equation}
where $E_{12}$ is the threshold energy for excitation from level 1 to level 2. 

In the case of a $\kappa$--distribution, the weighting with energy in the integral is quite different, \emph{c.f.} Equation  \eqref{e4}, and a knowledge of the behaviour of the collision strength at high energy becomes much more important. It is therefore necessary to use the raw (non-energy averaged) collision strengths.  While effective collision strengths have been published for almost all atomic species relevant to \HII regions and PNe, the raw collision strength data are much harder to find.  

For this work we have collated modern computed ``raw'' collision strength data for O~{\sc i}, N~{\sc ii}, O~{\sc iii}, S~{\sc iii}, and O~{\sc ii},  and older or limited data  for S~{\sc ii}, Ne~{\sc iii},  Ar~{\sc v}, Ne~{\sc iv}, Ar~{\sc iv}, and Ne~{\sc v}. We have no raw collision strength data for N~{\sc i}. Our data sources are listed in Table  \ref{t_1}.

\begin{table}[htdp]
\caption{Collision strength data sources}\label{t_1}
\begin{tabular}{llll}
\\
\hline
\hline
{Species} & {Authors} &{Reference} & {URL/source}\\  \hline
O {\sc iii} & Palay et~al. & 2012, MNRAS, 423, L35 & data from authors \\
O {\sc iii} & Aggarwal & 1993, ApJS, 85, 197 & data digitised from paper \\
O {\sc iii} & Aggarwal \& Keenan & 1999, ApJS, 123, 311& effective collision strengths only \\
O {\sc iii} & Lennon \& Burke & 1994, A\&AS, 103, 273 & TIPbase$^{1}$\\
S {\sc iii} & Hudson et~al. & 2012, ApJ, 750, 65 & data from authors \\
Ar {\sc iii} & Galav\'{i}s et~al. & 1995, A\&AS, 111, 347 & TIPbase\\
Ne {\sc iii} & Butler \& Zeippen & 1994, A\&AS 108, 1 & TIPbase\\
O {\sc ii} & Tayal & 2007, ApJS, 171, 331 & data from author\\
N {\sc ii} & Hudson \& Bell & 2004, MNRAS, 348, 1275 & APARC$^{2}$\\
N {\sc ii} & Tayal & 2011, \apjs, 195,12 & data from author \\
S {\sc ii} & Tayal \& Zatsarinny & \apjs, 2010,188, 32 & data from authors \\
O {\sc i} & Barklem & 2007, A\&A, 462, 781 & data from author \\
Ar {\sc iv} & Ramsbottom et~al. & 1997, MNRAS, 284, 754 & APARC\\
Ar {\sc v} & Galav\'{i}s et~al. & 1995, A\&AS, 111, 347 & TIPbase \\
 \hline
 \hline
\end{tabular}
\newline
$^{1}$The Iron Project database (TIPbase): http://cdsweb.u-strasbg.fr/tipbase\\
$^{2}$ APARC website: http://web.am.qub.ac.uk/apa\\
\end{table}%

\begin{figure}[htpb]
\includegraphics[scale=0.75]{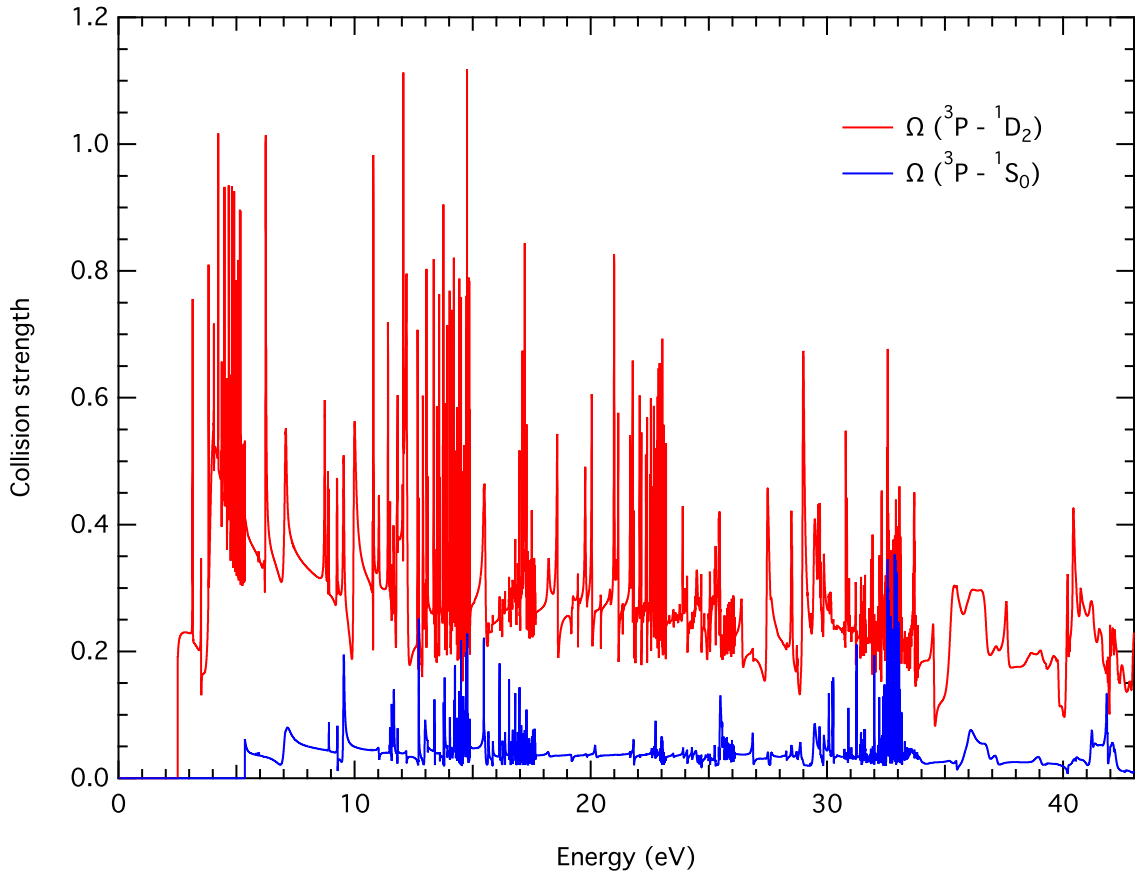}
\caption{Computed collision strength data for O~{\sc iii} from  \citetalias{Palay12}, shown here to 43 eV. Note the numerous resonances and edges, and the variation with energy  in the ${^3P} -  {^1D_2}$ transition.}\label{f_1}
\end{figure}

An example of the complexity of the raw collision strength data is shown for the $^1D_2$ and $^1S_0$ levels of O~{\sc iii} in Figure \ref{f_1}, where the data is taken from from \citet[hereafter, PNPE12]{Palay12}.  Note the numerous resonances and edges, and the systematic variation with energy  seen in the ${^3P} -  {^1D_2}$ transition.

The calculation of raw collision strengths is a very complex exercise, involving the coupling of many electrons, relativistic corrections, and a host of other computational issues. In general, there has been a steady improvement in the techniques of computation, so we need to be careful in using data from older sources. Given that an accurate knowledge of collision strengths is essential for determining electron temperatures and elemental abundances in nebulae, the errors that may be present in published data sets is a concern. In the following sub-sections we consider the possible effects of truncation of the energy range of the computed collision strengths, errors in the computed excitation energies, and absolute errors in the computed collision strengths on the collisional excitation rates.

\FloatBarrier 

\subsection{Errors in computed collision strengths}
Our knowledge of the absolute value of the collision strengths feeds directly into measurements of  electron temperatures and elemental abundances. Because of the complexity of calculating the collision strengths and the wide range of atomic species for which they are needed, these parameters are frequently only available at present from  a single source, if at all.  An exception to this is O~{\sc iii}, but even for this important species, they have only been computed four times in the past two decades, and only once in the past decade \citep{Aggarwal93, Lennon94, Aggarwal99,Palay12}.  Further, non-averaged collisions strengths (i.e., not convolved with M-B distributions) are difficult for the end user to obtain.  See Table \ref{t_1} above for details of the sources used.  

These computations vary considerably in their details, the upper energy limit of the computations (the truncation energy), and what physics is taken into account.  For O {\sc iii},  the most recent computations by \citetalias{Palay12} appear much the most reliable, as they take into account relativistic effects and have a much higher truncation energy (178.2eV, \emph{c.f.} 43.5eV for \citet{Aggarwal93} and 54.4eV for \citet{Lennon94}). For this reason the currently used values (see, e.g., \citet{Osterbrock06}) for calculating line flux ratios, and resultant electron temperatures, need to be revised, independently of any $\kappa$-distribution considerations.

\begin{figure}[htpb]
\includegraphics[scale=0.75]{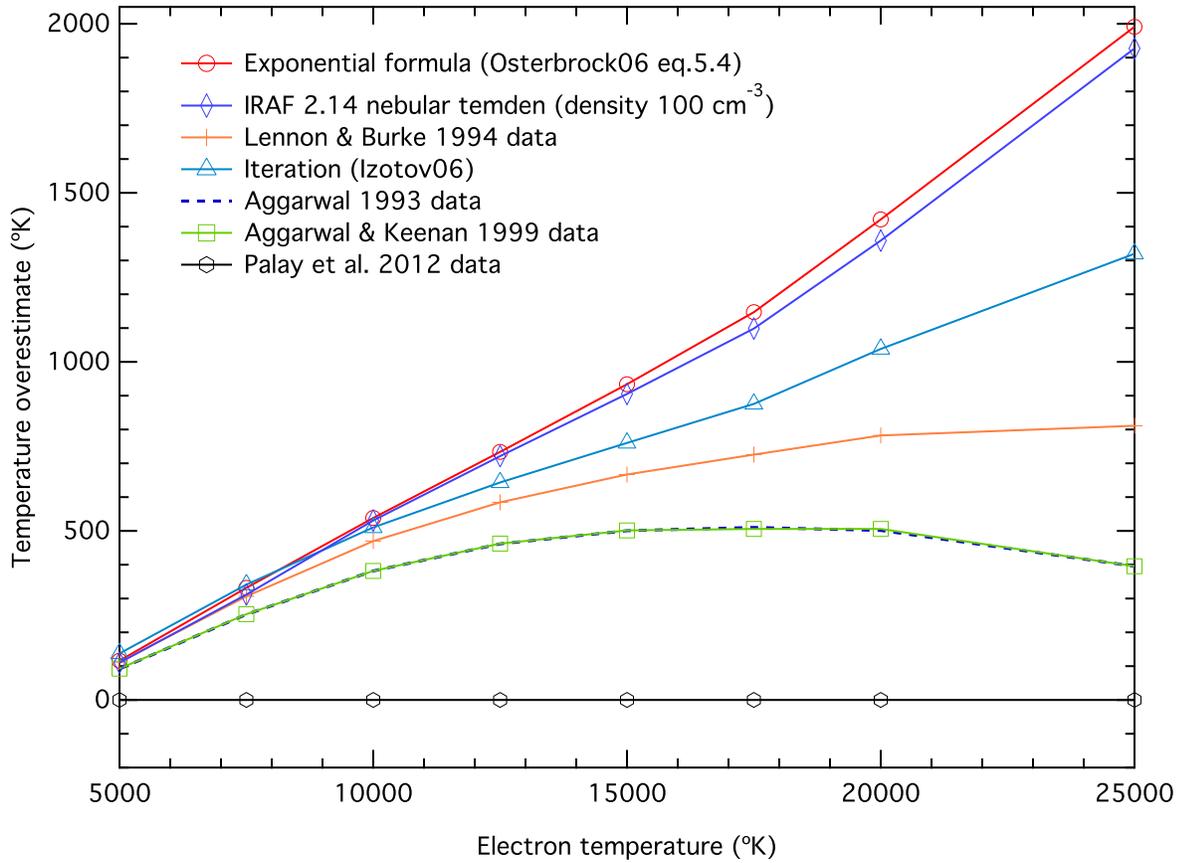}
\caption{Temperature excesses resulting from computing (M-B equilibrium) electron temperatures from [O~{\sc iii}] flux ratios, using older effective collision strength data, and approximate methods, compared to the results obtained using effective collision strengths derived from the latest data from \citetalias{Palay12}.}\label{f_2}
\end{figure}

We use the \citetalias{Palay12} data and detailed numerical integration as the baseline. This became available only after the finalization of our earlier paper.  The differences between these and earlier computations can lead to considerable differences in electron temperatures computed from CEL flux ratios, even for M-B equilibrium electron energy distributions.  Figure \ref{f_2} shows that use of the earlier data sources leads to systematic overestimates of [O {\sc iii}] electron temperatures for temperatures between 5,000 and 30,000K. The IRAF 2.14 results were obtained using the nebular/temden routine, which for the 11/2008 release adopts the \citet{Lennon94} effective collision strengths\footnote{PyNeb, a revised and extended Python-based version of the IRAF  nebular/temden routines has been developed \citep{Luridiana12} that uses more recent collision strength data than the older IRAF code. While it incorporates the O {\sc iii} data from \cite{Palay12}, this needs to be set as the default, manually.}. 

The overestimate of  [O {\sc iii}] electron temperatures implied by  Figure \ref{f_2} has a profound impact upon all previous abundance analyses of PNe and \HII regions, even before taking into account the effect of non-equilibrium $\kappa$ electron energy distributions. Wherever the $T_e$ + ionisation correction factor (ICF) method has been used, the overestimate in $T_e$ will result in a significant under-estimate in the chemical abundance.  The strong line techniques are also liable to revision, as the collision strength for the [O~{\sc iii}]  ${^3P} -  {^1D_2}$ transition is enhanced by about 30\% over the previous estimates. The effect on the strong line methods is discussed briefly in section 7, below, but these and other strong line effects will be the subject of a later paper.

\FloatBarrier 

\subsection{Errors in computed excitation energies}
Also critical to the accurate estimation of collision strength effects are errors in the computed threshold energies of the excited states. In some computations (e.g., \citet{Aggarwal93, Aggarwal99} for O {\sc iii}), there are non-trivial differences between the computed and the observed energies. \citetalias{Palay12} note that although their computed energies for O~{\sc iii} were quite close to the experimentally determined values, errors in effective collision strengths can arise from threshold energy discrepancies for low temperature excitations dominated by near-threshold resonances. They minimize these by adjusting the threshold energies to match the observed excitation energies. In the case of $\kappa$ distribution, where we integrate the raw collision strengths directly, it is essential that the threshold energies used in the integration ($E_{12}$ and $E_{13}$ in equation \eqref{e7}) correspond exactly to the values expressed in the collision strength data.  Using a threshold energy from a standard source that differs from the threshold indicated by the particular collision strength computations, can introduce errors in the excitation rate ratios, and, therefore, in the abundances determined assuming M-B equilibrium and the enhancement effects of a $\kappa$ distribution.

\subsection{Truncation of collision strength computations}
Finally, we need to consider the effect of truncating computations of collision strengths at high energies. Collision cross sections are calculated between the species excitation threshold energy and a computationally mandated upper limit. For the cross sections of O~{\sc iii} published in the past 20 years, this upper limit has ranged between 43.5eV \citep{Aggarwal93} and 178.2eV \citep{Palay12}.  Effective collisions strengths are computed by convolving the raw collision strengths with a M-B distribution, as in equation \eqref{e8}.  For temperatures typically found in \HII regions and PNe, the population in the M-B distribution  at high energies is sufficiently small that the truncation point for the raw collision strengths has little effect on the value of the effective collision strength.  However, $\kappa$ distributions can have significant populations at higher energies compared to the M-B, and the effect of truncating the collision strength computation can become much more apparent.

\begin{figure}[htpb]
\includegraphics[scale=0.75]{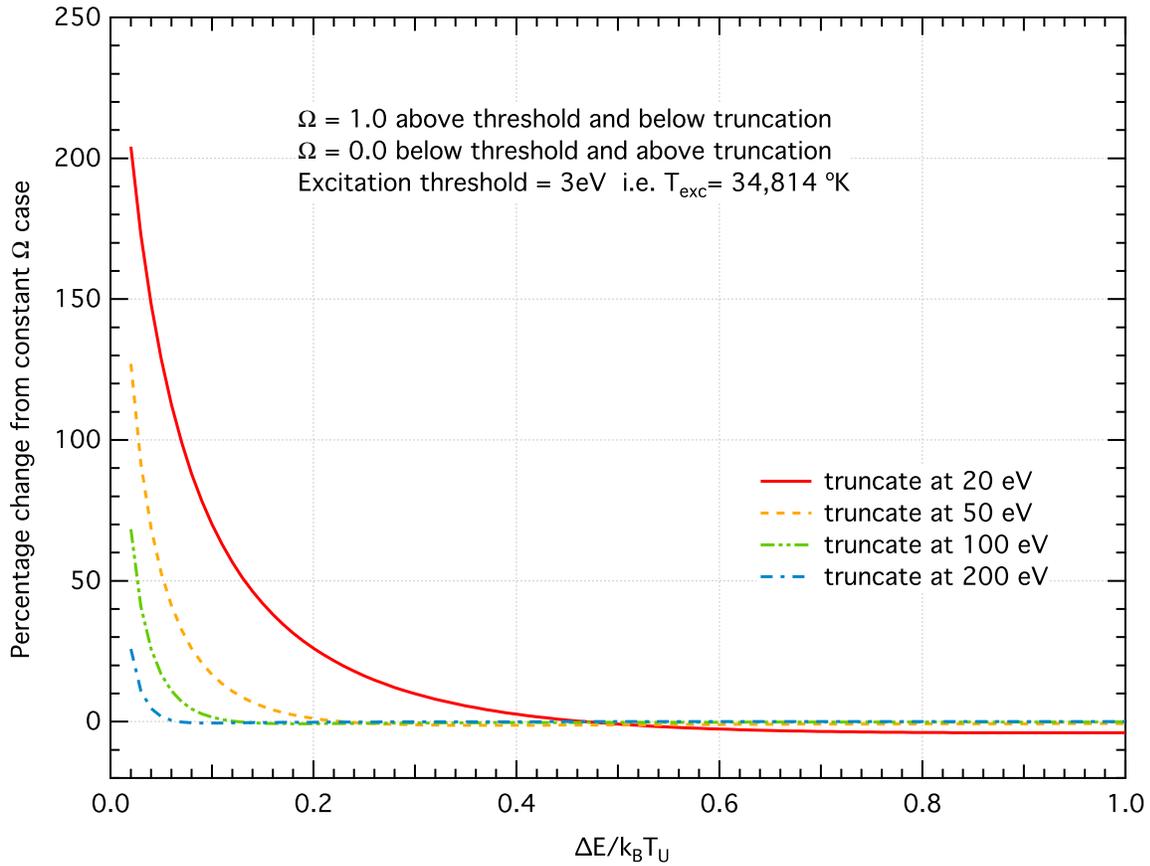}
\caption{Effect on the excitation rate of truncating the collision strength computations at a range of energies, for a $\kappa$=2 distribution.}\label{f_3}
\end{figure}

To demonstrate this effect, using an extreme case with $\kappa$=2, we adopt a simple model collision cross section: $\Omega$ = zero below the excitation threshold, $\Omega$ constant (=1) up to the truncation energy, and zero above that.  Specifying an excitation threshold energy  allows us to explore the effect of truncating the upper energy bound for the collision strength.  In this case we use 3.0eV, which sets the temperature of the point where  $\Delta E/ k_B T$ =1.0 to  $T_{exc}$=34,814K. We compare the computed truncated solution with the untruncated analytical solution, equation \eqref{e5}, in which $\Omega$ is constant to $\infty$. Figure \ref{f_3} shows the percentage difference between the computed values and the analytical value at low values of the parameter $\Delta E/ k_B T$ (i.e., at high temperatures),  truncating at 20, 50, 100 and 200eV. 

The effect is minor at low temperatures; for truncations above $\sim$50eV and temperatures typically found in \HII regions and PNe; and for  values of $\kappa \gtrsim$10.  In the EUV and in some supernova remnants, and for extreme values of $\kappa \gtrsim$ 1.5, the effect may need to be considered, both for $\kappa$ and M-B energy distributions.

\FloatBarrier 

\section{The Effect of $\kappa$ on recombination processes}
In this section, we examine first the effect of the $\kappa$-distribution on the recombination process. This links directly to the the shape of the the Bound-Free continuum which is used to determine recombination temperatures of H and He, and to the observed intensity of the recombination lines of heavy elements, which are used to determine chemical abundances.

\subsection{Recombination line effects}

One major consequence of adopting a $\kappa$ distribution for electron energies arises when comparing abundances determined using optical recombination lines (ORLs) and CELs. In the vast majority of \HII regions and PNe, the ORL abundance is systematically higher than the abundance derived from CEL measurements, the so called ``abundance discrepancy factor'', or ADF.  This has been known for decades and not satisfactorily explained \citep[see, e.g., ][]{Stasinska04}. As \citetalias{Nicholls12} have pointed out, the $\kappa$ distribution provides a simple and automatic explanation of the abundance ``discrepancy''. The reason for this can be understood by comparing the form of the $\kappa$ distribution to that of the M-B distribution.

The key characteristics of the $\kappa$ distribution, compared to a M-B distribution of the same internal energy, are that the peak of the distribution moves to lower energies; at intermediate energies there is a population deficit relative to the M-B distribution; and at higher energies the ``hot tail'' again provides a population excess over the M-B.   \citepalias[See Figures 1-3 of][]{Nicholls12}.  The $\kappa$ distribution behaves as a M-B distribution at a lower peak temperature, but with a significant  high energy excess. 

\begin{figure}[htpb]
\includegraphics[scale=0.75]{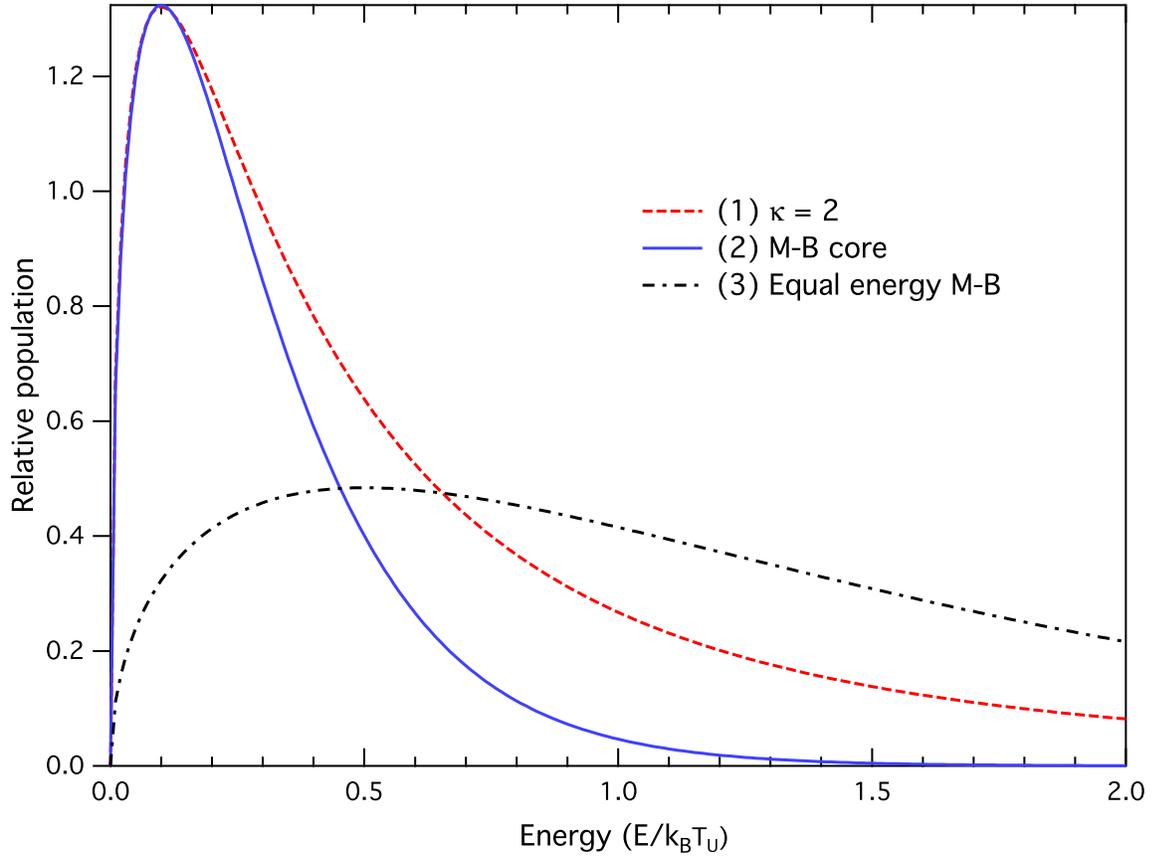}
\caption{(1) red dashed curve: $\kappa$ = 2 distribution; (2) blue curve: ``core'' M-B distribution fitted to the $\kappa$ peak; (3) black dash/dot curve: M-B distribution with the same internal energy as the $\kappa$. The areas under the red (dashed) and black (dot-dashed) curves (i.e., the total electron densities) are equal, and greater than the area under the blue curve.}\label{f_4}
\end{figure}

The two distributions peak at different values of the energy, $E$.  The peak of the Maxwell distribution (for the energy form of the distribution) is at $E = \frac{1}{2}k_BT_U$.  For the $\kappa$ distribution, the peak occurs at $\frac{1}{2}k_BT_U (2\kappa - 3)/(2\kappa + 1)$ \citepalias{Nicholls12}. Thus, for all valid values of $\kappa$ ($\frac{3}{2} < \kappa < \infty$), the $\kappa$ distribution peaks at a lower energy than the M-B.  This is illustrated in Figure \ref{f_4}, for $\kappa$ = 2.

\FloatBarrier 
For recombination, or any other physical process that is primarily sensitive to the low energy electrons, the critical point to note is that the form of the $\kappa$ distribution at lower energies (up to and just past the peak energy) is very similar indeed to a M-B distribution.  This is shown in Figure \ref{f_4}, where a M-B distribution (blue solid curve) has been peak-fitted to a $\kappa$ = 2 distribution (red, dashed curve), adjusting the M-B temperature to $T_{core} = T_U (1- \frac{3}{2\kappa})$ and matching peak heights.  The total area under the M-B ``core'' is less than the area under the $\kappa$ curve.

For any physical process that involves mainly the low energy electrons, such as recombination line emissions, reactions ``see'' the cool M-B core distribution.   In other words, any physical property sensitive to the region of the electron energy distribution around or below the distribution peak will interact with a $\kappa$ electron energy distribution as if it were a M-B distribution at a lower temperature than the M-B with the same kinetic temperature and electron density as the $\kappa$ distribution, and with a slightly lower total internal energy than the $\kappa$-distribution.

So how does this impact on recombination line abundances and temperatures?  In order of importance, the first effect of the kappa distribution on ORLs is the difference between the apparent temperature of the low energy part of the energy distribution that is most important in determining the intensities of the recombination lines, compared to the true internal energy temperature.  The second effect arises from the population of electrons in the energy peak of a kappa distribution, compared to the total population. The third effect is the slight difference in shape between the peak of a kappa distribution and the best fit M-B distribution.

\subsection{Correcting the recombination temperature}

First, the most obvious effect of a kappa distribution is that it shifts the peak of the energy distribution to lower energies, compared to a M-B distribution with the same kinetic temperature.  The rate of recombination rate falls off strongly with increasing energy---for hydrogen below the phoionization threshold, the  recombination rate depends on $\nu^{-3}$ \citep[e.g., ][]{Osterbrock06}. This means that the low energy electrons play the dominant role in recombination processes.  Recombination processes experience the $\kappa$ distribution as a M-B distribution at a temperature $T_{core}$.  Thus, in using recombination temperatures in the presence of $\kappa$ distributions to estimate the kinetic or internal energy temperature, $T_U$, we need to increase the apparent recombination line temperature by a factor:

\begin{equation}\label{e9}
 T_U/T_{core} = \kappa/(\kappa-3/2)\ . 
 \end{equation}
 The difference between the distributions is visually slight for higher kappa values (smaller deviation from thermal equilibrium), but even minor deviations from equilibrium can be sufficient to explain the ``abundance discrepancy factor''. 

\subsection{Correcting the electron density}

Second, we need to apply a correction to the apparent electron density. The reason for this is that a M-B distribution at a temperature $T_{core}$ and with the $\it{same}$ total energy as the $\kappa$ distribution with a kinetic temperature $T_U$ will have a peak at a $\it{higher}$ value of $n(E)$ than the $\kappa$.  To fit the M-B distribution to the $\kappa$---in other words, to simulate what recombination processes react to when they meet a $\kappa$ distribution---it is necessary to reduce the total electron density by a factor that depends on $\kappa$.

We can calculate the electron density correction analytically by equating the peak of the M-B electron energy distribution n(E) at a temperature $T_{core}$ to the peak value of the $\kappa$ distribution at a temperature $T_U$. It is relatively straight forward to show that the effective (apparent) electron density, $N_e$(eff) is related to the actual electron density $N_e$, by:

\begin{equation}\label{e10}
\frac{N_e(\text{eff})}{N_e}=\frac{((\kappa+1)/(\kappa+\frac{1}{2}))^{\kappa+1} \sqrt{(\kappa+\frac{1}{2})} (\kappa-\frac{3}{2}) \Gamma(\kappa-\frac{1}{2})}{\Gamma(\kappa+1) \sqrt e}\ .
\end{equation}

\begin{figure}[htpb]
\includegraphics[scale=0.75]{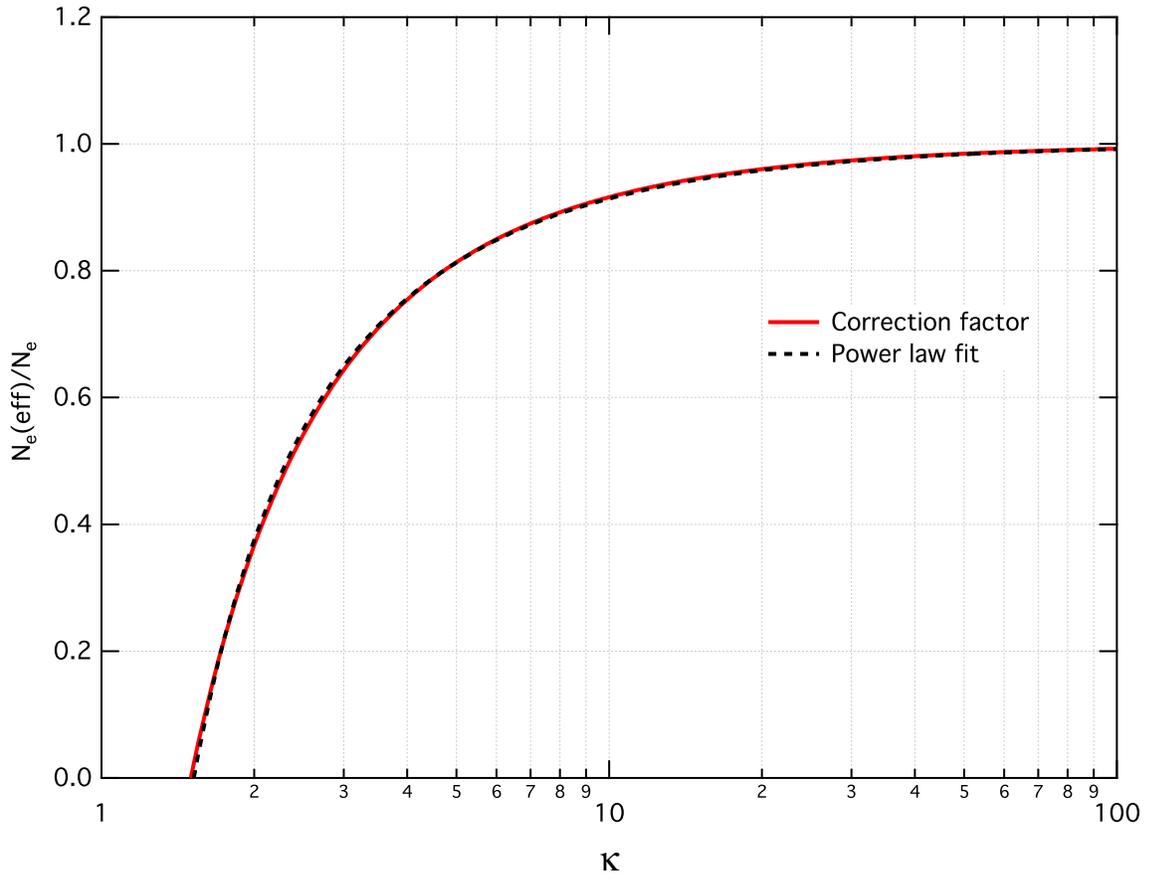}
\caption{Electron density correction as a function of $\kappa$}\label{f_5}
\end{figure}

For values of $\kappa \gtrsim $ 10, this factor is close to unity, and in most conditions likely to be found in \HII regions and PNe \citepalias{Nicholls12} is unlikely to substantially affect the physics.  The correction factor is shown in Figure \ref{f_5} as a function of $\kappa$. The recombination process ``sees'' a lower electron density for all values of $\kappa$, but for typical values $\sim$10, the difference between effective and true electron densities  is less than 10\%.

For computational purposes, the curve can be fitted with a simple power law (reciprocal), also shown in Figure \ref{f_5}:
\begin{equation}\label{e11}
\frac{N_e(\text{eff})}{N_e}= 1.0 - 0.8/(\kappa - 0.72)\ .
\end{equation}

\FloatBarrier 

\subsection{Correcting the low energy shape of the distribution}

\begin{figure}[htpb]
\includegraphics[scale=0.75]{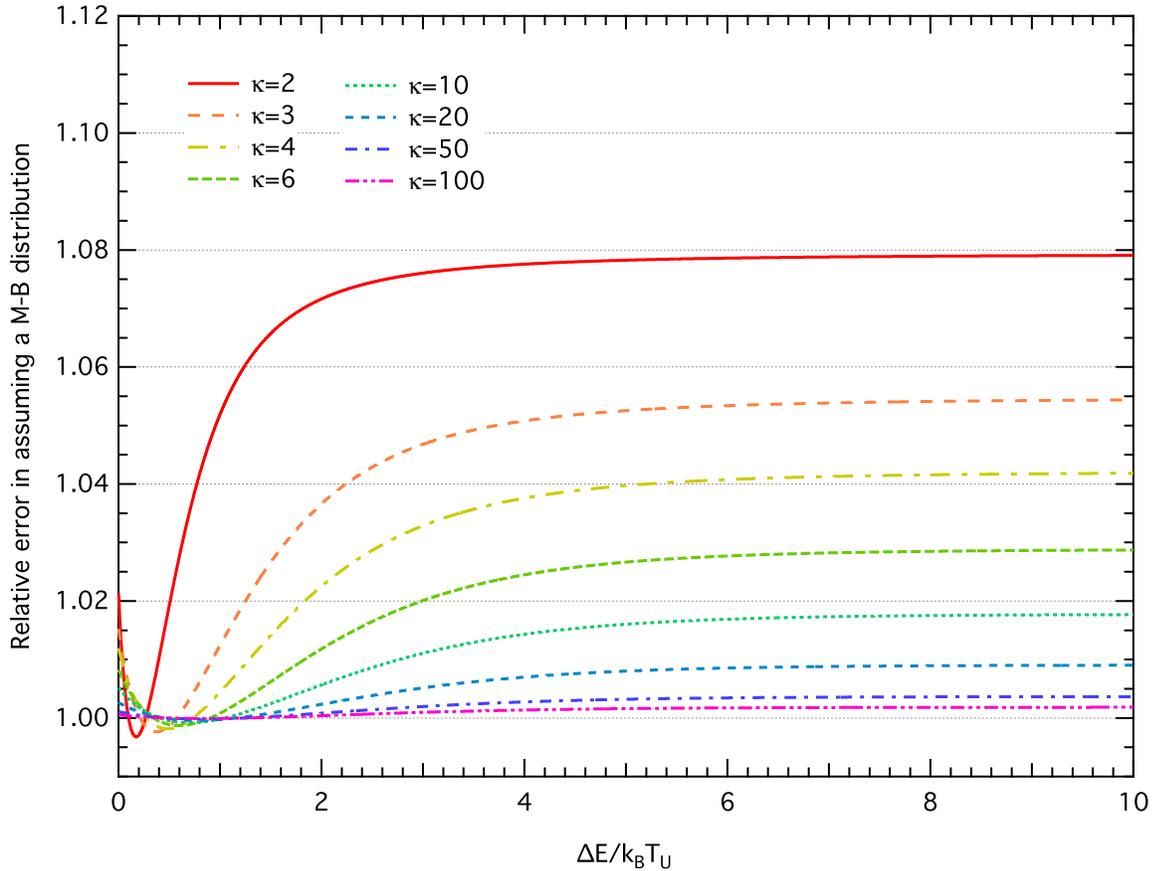}
\caption{Error in assuming a fitted M-B distribution instead of a $\kappa$ distribution, as a function of $\Delta E/k_BT_U$ and $\kappa$}\label{f_6}
\end{figure}

The third effect is that the shape of the ``fitted'' M-B distribution differs slightly from the peak of the $\kappa$ distribution. Figure \ref{f_6} shows the difference in  recombination electrons as a function of $\Delta E/k_BT_U$, using a weighting factor of 1/E to account for a typical energy dependence of the recombination process, and normalised so that the total number of electrons at the distribution peaks are the same. It shows that for a typical value of $\kappa$ of 10, the difference in the $\kappa$ distribution and the fitted M-B leads to an error of less than 2\%.

\FloatBarrier 
\subsection{Effect of $\kappa$ on recombination rates}
The recombination rate (in $s^{-1} cm^{-3}$) for hydrogen ions combining with electrons is given by  $N_e N_p \alpha$, where $N_e$ and $N_p$ are the densities of electrons and protons and $\alpha$ is the recombination rate, which for an electron energy distribution $f(E)dE$ is given by:
\begin{equation}\label{e12}
\alpha =  \int\limits_0^\infty \sqrt{\frac{2}{m_e}}\sqrt{E}\ \sigma (E) f(E)  dE\ ,
\end{equation}
where $\sigma (E)$ is the recombination cross section. It is related via the Milne Relation to the ionization cross section $a_\nu$ by:
\begin{equation}\label{e13}
\sigma(E) = \frac{g_1}{g_2} \frac{2 h^2 \nu ^2}{m_e c^2} \frac{1}{E} a_\nu \ ,
\end{equation}
where $g_{1,2}$ are the statistical weights of the lower and upper levels, $h$ is the Planck constant, $m_e$ is the electron mass,  $c$ is the speed of light and $\nu$ is the photon energy above the threshold (expressed as a frequency).

For hydrogen, $a_\nu$ can be expressed approximately as:
\begin{equation}\label{e14}
a_\nu = a_T \left(\frac{\nu}{\nu_T}\right)^{-3} \ ,
\end{equation}
where $a_T$ is the threshold value of the ionization cross section and $\nu_T$ is the threshold frequency.

Inserting these values into equation \eqref{e12} and gathering the energy-independent components outside the integral we get:
\begin{equation}\label{e15}
\alpha =   \sqrt{\frac{2}{m_e}}\frac{g_1}{g_2} \frac{2 h^2 \nu ^2}{m_e c^2}\ a_T\left(\frac{\nu}{\nu_T}\right)^{-3} \int\limits_0^\infty \frac{1}{\sqrt{E}}\ f(E)  dE
\end{equation}

We can calculate the ratio of the recombination rates for a $\kappa$ distribution to a M-B distribution by substituting the appropriate forms for $f(E)$:
\begin{equation}\label{e16}
\frac{\alpha _\kappa}{\alpha_{\textrm{M-B}}} = \int\limits_0^\infty \frac{1}{\sqrt{E}}\ f_\kappa (E)  dE \bigg / \int\limits_0^\infty \frac{1}{\sqrt{E}}\ f_{\textrm{M-B}} (E)  dE
\end{equation}

This simplifies to a form similar to the analytical expression for collisional excitation with a constant collision strength from equation \eqref{e5}, but  in this case with $E_{12}=0$:
\begin{equation}\label{e17}
\frac{\alpha _\kappa}{\alpha_{\textrm{M-B}}} =  \frac{\Gamma(\kappa+1)}{(\kappa-\frac{3}{2})^{3/2}\Gamma(\kappa-\frac{1}{2})}\left(1-\frac{3}{2\kappa} \right)
\end{equation}

This implies the hydrogen ion recombination rates are enhanced, but for a typical value, $\kappa$=10, only by 4.3\%. Typical values for the recombination rate ratios are given in Table \ref{t_2}:

\begin{table}[htbp]
\caption{Recombination rate ratios as a function of $\kappa$}
\begin{center}
\begin{tabular}{lllllllll}
\hline 
\hline \\
$\kappa$ & 2 & 3 & 4 & 6 & 10 & 20 & 50 & 100 \\
$\alpha_\kappa /\alpha_{\textrm{M-B}}$ & 1.59577 & 1.22842 & 1.14184 & 1.08073 & 1.04338 & 1.02011 & 1.00771 & 1.0038 \\
\\ \hline
\hline \\
\end{tabular}
\end{center}
\label{t_2}
\end{table}
\subsection{Recombination lines: summary}
In summary, when interpreting a $\kappa$ distribution as if it were a M-B distribution: (1) apparent recombination temperatures need to be increased by a factor $\kappa/(\kappa - 3/2)$; (2) apparent electron densities need to be divided by the correction factor in equation \eqref{e10}, to get the true electron densities and kinetic temperatures; (3) the ``shape'' correction is sufficiently small that it can be neglected; and (4) recombination rates are slightly enhanced, as per Table \ref{t_2} and equation \eqref{e17}.  Note that the corrections to the recombination rate are only  applicable to recombination of ions with recombination coefficients similar to Hydrogen.

\section{Collisionally Excited Lines}

\subsection{Effect on CEL Intensities}

\begin{figure}[htpb]
\includegraphics[scale=0.75]{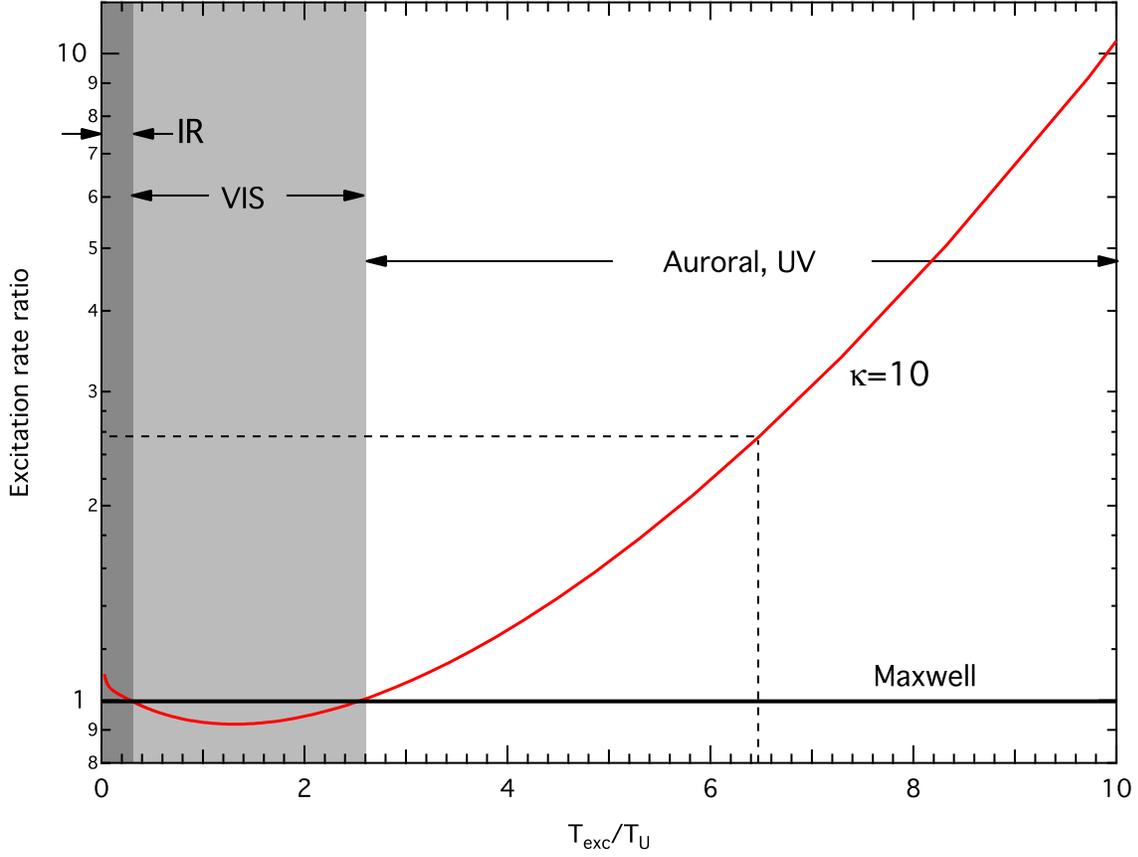}
\caption{The collisional excitation rate for $\kappa$=10  compared to a M-B distribution, plotted as a function of the excitation threshold energy (expressed as an equivalent temperature) divided by the kinetic temperature $T_U$. Setting the kinetic temperature $T_U$ to a typical nebular temperature of $10^4$K allows us to locate the excitation temperature of the O~{\sc iii} $^1S_0$ level. It is marked by the vertical dashed line. Where this intercepts the  $\kappa$ curve shows the enhancement of the excitation rate ratio (and therefore, of the population in that level, relative to the M-B population). This illustrates the generic behaviour of all CELs. When $T_{exc}/T_{U}$ is low, such as for transitions in the IR and FIR, the emission line intensities are slightly enhanced (dark grey area). In the central (light-gray) region typical of transitions giving rise to lines at optical wavelengths, a mild reduction in line intensity is expected.  For $T_{exc}/T_{U} \gtrsim 3$, appropriate UV or ``auroral'' line intensities are either enhanced or strongly enhanced. }\label{f_7}
\end{figure}

In Figure \ref{f_7} we show, for $\kappa = 10$ and a kinetic (internal energy) temperature $T_U$=10,000K,  the relative collisional excitation rate relative to a M-B distribution as a function of $T_{exc}/T_{U}$ for the [O~{\sc iii}] $\lambda$ 4363 auroral line, computed using the detailed collisions strengths for O~{\sc iii} from \citetalias{Palay12}.  Any other CEL would produce a similar curve, so Figure  \ref{f_7} provides a generic description of the effects of a $\kappa$-distribution on CEL intensities. Note that for a fixed kinetic temperature $T_U$, positions along the x-axis correspond to values of the CEL excitation temperature in units of $10^4$K.  The axis could equally well be looked at by scaling the kinetic temperature for a fixed excitation temperature, but here we want to differentiate the effects of $\kappa$ on lines with different excitation temperatures at a fixed kinetic temperature.

So what does this mean for different atomic species, energy levels and radiative transitions?  Figure \ref{f_7} can be divided into three parts, marked in different shades.  The left-most dark gray segment corresponds to fine-structure levels with low excitation energy. These typically correspond to far infrared lines.  For such levels, the population rate is slightly enhanced, leading to slightly higher line fluxes.  

The middle section (mid-gray) corresponds to the excitation of  the strong visible transitions, with excitation energies of a few eV.  An example would the [S~{\sc ii}] lines at 6731\AA\ and 6716\AA, with excitation temperatures of $\sim$21,400K (for $T_U$=10,000K, this corresponds to $x$=2.14 in Figure \ref{f_7}). The collisional excitation rates for these lines are mildly reduced in a $\kappa$-distribution compared to a M-B distribution.

The third, right-most section shows the excitation energies where the population rate will be enhanced or strongly enhanced by the $\kappa$ distribution, compared to the M-B.  This region is appropriate to either highly-excited UV lines, or the ``auroral'' lines in the visible spectrum.  Examples include the [O~{\sc iii}] UV lines at 2321, 2331\AA\ and the auroral line at 4363\AA, with an excitation temperature of $\sim$62,000K corresponding to $x$=6.2 in Figure \ref{f_7}.

In summary then, for a $\kappa$ distribution the far-IR transitions are slightly enhanced, and the strong emission lines used in the optical to obtain CEL abundances will be mildly reduced. However, we expect the UV lines, such as the important  C~{\sc ii} or C~{\sc iii} intercombination lines,  to be strongly enhanced, and the ``auroral'' lines used in temperature diagnostics also to show strong enhancements in more metal-rich \HII regions.

The relative effect of $\kappa$ at different metallicities is interesting to consider.  Plasmas with higher metallicities cool faster than plasmas with low metallicities.  If we set the kinetic temperature for Figure \ref{f_7} to 20,000K, i.e., to a lower metallicity, the excitation temperatures are now scaled in units of $2\times10^4$K. Thus for the O~{\sc iii} $^1S_0$ level, the excitation temperature occurs at $x \sim 3.1$, and at this point on the curve, the excitation enhancement by the $\kappa$ distribution is much lower, $\sim$1.1 \emph{c.f.} $\sim$2.6.  The precise effect on the line flux ratio used to measure the electron temperature depends as well on the relative enhancement of the 5007\AA\  and 4959\AA\ lines, which will also fall with lower metallicities.  The process is not simple because of the interconnected effects, and is best explored with photoionization models that take the $\kappa$ effects into account.  We have extensively updated the MAPPINGS photoionization code to take into account both the $\kappa$ effects and the latest atomic data.  We explore these effects in a subsequent paper \citep{Dopita13} using this code.

In the following section we explore the explicit effects of the $\kappa$ distribution on line flux ratios.

\subsection{Temperature-sensitive line ratios}

Collisionally excited line ratios are central to the measurement of electron temperatures in \HII regions and PNe.  Most frequently, the ratio of optical forbidden lines of O~{\sc iii} at 5007\AA, 4959\AA\ to the ``auroral'' transition at 4363\AA\ has been employed. However, many others can be used when bright lines are observed, such as the [N~{\sc ii}], [O~{\sc ii}], [S~{\sc ii}], [S~{\sc iii}], [Ar~{\sc iii}], [Ar~{\sc iv}], [Ar~{\sc v}], or [Cl~{\sc iii}] forbidden line ratios \citep[see, e.g.,][]{Peimbert03}. The measurement of electron temperatures depends on having two well-separated excited fine-structure energy levels for which an equation of the form of Equation \eqref{e6} or \eqref{e7} applies. An idealised three energy level arrangement is shown in Figure \ref{f_9}(a), which illustrates the transitions involved in the formation of temperature-sensitive line ratios.

\begin{figure}[htpb]
\includegraphics[scale=0.75]{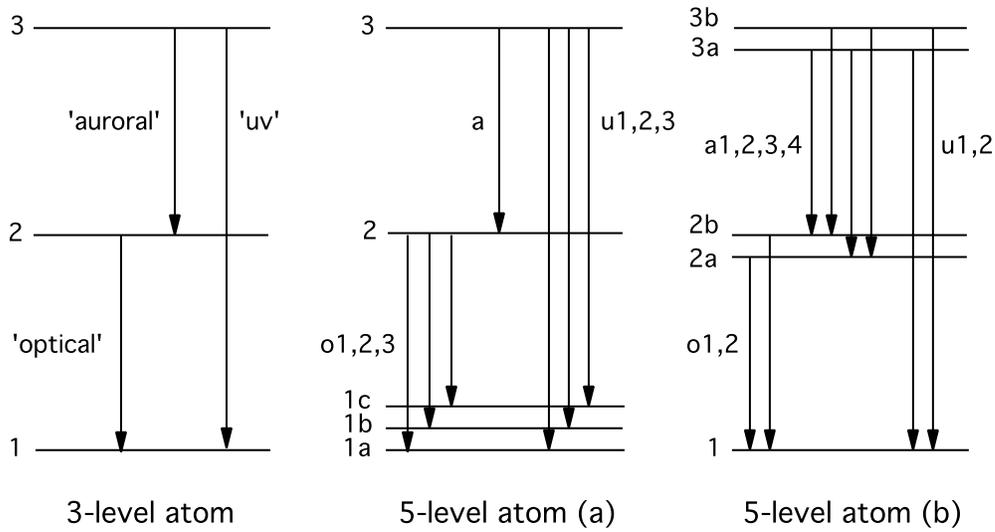}
\caption{Energy level arrangements for temperature-sensitive line ratios. The left diagram illustrates the simplified 3-level arrangement illustrating the lines involved. The center diagram represents the level configuration for $p^2$ and $p^4$ ions, and the right hand panel, $p^3$. The doublet and triplet spacings are not drawn to scale. Some lines arise from doubly-forbidden transitions with very low transition probabilities.}\label{f_8}
\end{figure}

Among the species actually employed to measure electron temperatures, there are two principal energy level structures.  The first of these  are the $p^2$ ions such as O~{\sc iii}, and the $p^4$ ions  such as O~{\sc i}, which have a very similar fine-structure level configuration, as shown in the the second panel of Figure \ref{f_8} (case a). The second group consists of the $p^3$ ions, such as O~{\sc ii}, which has a doublet structure in the excited states as shown in the third panel of Figure \ref{f_8} (case b). These ions are most frequently used to determine electron densities since the closely spaced excited states have different transition probabilities, and undergo collision de-excitation at different densities. 

The $p^2$ and $p^4$ ions have a triplet ground state ($^3P_0, ^3P_1, ^3P_2$) and singlet upper states, $^1D_2$ (lower) and $^1S_0$ (upper).  Examples  include N~{\sc ii}, O~{\sc iii}, S~{\sc iii}, Ne~{\sc v} and  Ar~{\sc v} ($p^2$ configuration) and O~{\sc i}, Ne~{\sc iii}, and Ar~{\sc iii} ($p^4$ configuration).  The $p^3$ ions have a single ground state (usually $^4S^0_{3/2}$) and a pair of closely spaced doublet upper states, usually $^2D^0_{3/2}, ^2D^0_{5/2}$ (lower) and $^2P^0_{1/2}, ^2P^0_{3/2}$ (upper). Examples of this form include N~{\sc i}, O~{\sc ii}, S~{\sc ii}, Ar~{\sc iv}, and Ne~{\sc iv}.   

To calculate the flux ratio, we must take into account the branching ratio for transitions from the uppermost state, the summed transition probabilities for transitions to multiple ground states, and the transition-probability-averaged energies for the multiple optical lines.  For the $p^2$ or $p^4$ ions  (case a), the general expression for the ratio of flux of the auroral line from level 3 to 2 to the fluxes of the optical lines from level 2 to 1b and 2 to 1c (ignoring the doubly forbidden line from level 2 to 1a) is given by the a generalised inverse of equation 5.1 in \citet{Osterbrock06}:
\begin{equation}\label{e18}
\frac{j_{\lambda a}}{j_{\lambda o2}+j_{\lambda o3}}=\frac{\overline{\Upsilon}_{13}}{\overline{\Upsilon}_{12}}\left[\frac{A_a}{A_a+\Sigma A_u}\right]\frac{(A_{o2}+A_{o3})\nu(\lambda_a)}{A_{o2}\nu(\lambda_{o2}) + A_{o3}\nu(\lambda_{o3})}\ \exp\left[\frac{-E_{23}}{k_B T}\right]\ ,
\end{equation}
where $E_{23}$ is the energy gap between the two singlet states,  $\overline{\Upsilon}_{12}$ and $\overline{\Upsilon}_{13}$ are the (mean) effective collision strengths for collisional excitation from the triplet ground states to the lower and upper singlet states, given by (e.g.):
\begin{equation}\label{e19}
\overline{\Upsilon}_{12}= \frac{\Upsilon_{1a\rightarrow 2}g_{1a} +\Upsilon_{1b\rightarrow 2}g_{1b} + \Upsilon_{1c\rightarrow 2}g_{1c}}{g_{1a}+g_{1b}+g_{1c}} \ ,
\end{equation}
and where $g_{1a}$, $g_{1b}$ and $g_{1c}$ are the statistical weights of levels 1a, 1b, and 1c; and $\Sigma A_u$ is the total transition probablility for the transitions between the upper singlet state (3) and the triplet ground states (1a, 1b and 1c), 
\begin{equation}\label{e20}
\Sigma A_u=  A_{u1}+A_{u2}+A_{u3}\ .
\end{equation}

In practice, one of the transitions from the singlet upper state to one of the triplet ground states is doubly forbidden and its transition probability is negligible. The term in equation $\eqref{e18}$ in the square brackets is the branching ratio, i.e., the fraction of atoms excited to level 3 that decay to level 2, and the term following that is the energy weighting for the transition probabilities.

For the $p^3$ ions, the expression for the flux ratio is similar to equation $\eqref{e18}$:
\begin{equation}\label{e21} 
\frac{\Sigma j_{\lambda a}}{j_{\lambda o1}+j_{\lambda o2}}=\frac{{\Upsilon}_{13}}{{\Upsilon}_{12}}\left[\frac{\Sigma A_a}{\Sigma A_a+\Sigma A_u}\right]\frac{(A_{o1}+A_{o2})\overline{\nu_a}}{A_{o1}\nu_{o1}+A_{o2}\nu_{o2}}\ \exp\left[\frac{-E_{23}}{k_B T}\right]\ ,
\end{equation}
where
\begin{equation}\label{e22}
\overline{\nu_a}=\left(\frac{ A_{a1}\nu_{a1}+A_{a2}\nu_{a2}+A_{a3}\nu_{a3}+A_{a4}\nu_{a4}}{\Sigma A_a}\right)\ ,
\end{equation}
and
\begin{equation}\label{e23}
\Sigma A_a= A_{a1}+A_{a2}+A_{a3}+A_{a4}\ ,
\end{equation}
and
\begin{equation}\label{e24}
\Upsilon_{12} = \Upsilon_{1\rightarrow 2a} + \Upsilon_{1\rightarrow 2b}\ ,
\end{equation}
with $\Sigma A_u$ and $\Upsilon_{13}$ defined analogously.

For each of the ions we consider here, the values of the various constants entering in these equations are listed in the tables in the Appendix. As shown in Figure \ref{f_8}, the transitions `o1' and `o2' are the ``optical'' transitions, from the two middle levels to the ground state; transitions `a1', `a2', `a3' and `a4' are the four ``auroral'' lines from each of the upper levels to each of the middle levels; and transitions `u1' and `u2' are the ``UV lines'' from the upper two levels to the ground state (frequently in the optical, not the UV, spectrum).  $\Sigma j_{\lambda a}$ is the total flux of the (four) auroral transitions, $j_{\lambda o1}$ and $j_{\lambda o2}$ are the fluxes of the two optical lines.

In some cases where wavelengths of the auroral lines are not well placed, it is more convenient to use the UV lines in combination with the optical lines to measure temperature dependent flux ratios.  Examples where this is used in the IRAF/temden routine are S~{\sc ii}, Ne~{\sc iv} and Ar~{\sc iv}.  However, in principle, UV lines can be used equivalently to auroral lines. This can be useful at higher redshifts.  As the UV and auroral lines both originate from the uppermost of the levels (3 or 3a, 3b in Figure \ref{f_8}), their relative fluxes are related via the branching ratio and the energies of the transitions. This may be expressed in general as a ratio:
\begin{equation}\label{e25}
\frac{\text{flux(UV)}}{\text{flux(auroral)}} = \frac{\displaystyle\sum\limits_i(A_{ui}/\lambda_{ui})}{\displaystyle\sum\limits_i(A_{ai}/\lambda_{ai})}
\end{equation}
However, most of the $p^3$ ions are also strongly density sensitive, so flux ratios using these lines---auroral or UV---will only give useful temperatures at densities  $\lesssim$5 cm$^{-3}$.

\subsection{Excitation rate ratios}
The generalised version of Equations \eqref{e6} and \eqref{e7} for the energy-dependent $\Omega$ case are:
\begin{equation}\label{e26}
\frac{R_{13}}{R_{12}}=\frac{\int\limits_{E_{13}}^\infty \Omega_{13}(E)\ \mathrm{ exp} \left[-E/k_BT_U\right] dE}{\int\limits_{E_{12}}^\infty \Omega_{12}(E)\ \mathrm{ exp} \left[-E/k_BT_U\right] dE}
\end{equation}
for the M-B electron distribution and
\begin{equation}\label{e27}
\frac{R_{13}}{R_{12}}=\frac{\int\limits_{E_{13}}^\infty \Omega_{13}(E) / [1 + E/((\kappa-\frac{3}{2}) k_BT_U)]^{\kappa + 1} dE}{\int\limits_{E_{12}}^\infty \Omega_{12}(E) / [1 + E/((\kappa-\frac{3}{2}) k_BT_U)]^{\kappa + 1} dE} ,
\end{equation}
for the $\kappa$-distribution. 

We can now generalize the expression for the flux ratio for variable $\Omega$s, using equations $\eqref{e18}$ and $\eqref{e27}$:
\begin{equation}\label{e28}
\frac{j_{\lambda a}}{j_{\lambda o1}+j_{\lambda o2}}=f_1(A,\lambda)\ \frac{\int\limits_{E_{13}}^\infty \overline{\Omega}_{13}(E) / [1 + E/((\kappa-\frac{3}{2}) k_BT_U)]^{\kappa + 1} dE}{\int\limits_{E_{12}}^\infty \overline{\Omega}_{12}(E) / [1 + E/((\kappa-\frac{3}{2}) k_BT_U)]^{\kappa + 1} dE}\ ,
\end{equation}
where
\begin{equation}\label{e29}
f_1(A,\lambda)=\left[\frac{A_a}{A_a+A_{u1}+A_{u2}}\right]\frac{(A_{o1}+A_{o2})\nu(\lambda_{a})}{A_{o1}\nu(\lambda_{o1}) + A_{o2}\nu(\lambda_{o2})} \ ,
\end{equation}
and $\overline{\Omega}_{13}, \overline{\Omega}_{12}$ are the statistical weight averaged $\Omega$s, defined analogously to equation $\eqref{e19}$.  This equation allows us to calculate the line ratios for any of the relevant atomic species and for any value of $\kappa$.  The values of the parameter $f_1(A,\lambda)$ for several atomic species are given in Table \ref{t_7} in the Appendix.

Similarly, equation \eqref{e21} can be generalized for non-M-B populations for the $p^3$ ions as:
\begin{equation}\label{e30}
\frac{j_{\lambda a}}{j_{\lambda o1}+j_{\lambda o2}}=f_2(A,\lambda)\ \frac{\int\limits_{E_{13}}^\infty \overline{\Omega}_{13}(E) / [1 + E/((\kappa-\frac{3}{2}) k_BT_U)]^{\kappa + 1} dE}{\int\limits_{E_{12}}^\infty \overline{\Omega}_{12}(E) / [1 + E/((\kappa-\frac{3}{2}) k_BT_U)]^{\kappa + 1} dE}\ .
\end{equation}
where
\begin{equation}\label{e31}
f_2(A,\lambda)=\left[\frac{\Sigma A_a}{\Sigma A_a+\Sigma A_u}\right]\frac{(A_{o1}+A_{o2})\overline{\nu_{a}}}{A_{o1}\nu_{o1}+A_{o2}\nu_{o2}}\ .
\end{equation}
The values of the parameters  for the $p^4$ ions are also given in Table \ref{t_8} in the Appendix.

Tables \ref{t_7} and \ref{t_8} also show the values of $f_1(A,\lambda)$ and $f_2(A,\lambda)$ using the UV lines instead of the auroral lines.

\subsection{Plotting the temperature-sensitive line ratios}

The simplest way to determine the electron temperature $T_e$ from the line ratios is to use the IRAF/SCSDS/nebular/temden routine \citep{Shaw95}, or the more recent PyNeb code \citep{Luridiana12}.  Alternatively, one can use \citet[Figure 5.1]{Osterbrock06}, reading off the temperature from the line ratio graph, using the inverse of equation \eqref{e12} above.  This does not take into account that the collision strengths (and even the effective collision strengths) are not constant with temperature, but in general are complex functions of the energy above the threshold (see  Figure \ref{f_1} below).  For a M-B distribution, one can use the effective collision strengths for each temperature, leading to a more accurate function of line ratio  vs. temperature.  A further improvement to this process was used by \citet{Izotov06} who derived an iterative formula to obtain $T_e$ from the line ratio measurements.

However, current methods only apply where there is thermal equilibrium, and in the non-equilibrium $\kappa$ distribution case, it is necessary to calculate the integrals in equation \eqref{e21} numerically, using the original collision strength data (not the thermally averaged values).  This leads to a graph similar to that presented in \citet{Osterbrock06}, with a series of curves for each value of $\kappa$ required.  The result is simple to determine.

As noted for equation \eqref{e18}, in this paper (except in section 6.1) we break with tradition and invert the equation, as it is easier to understand the correlation between an increasing upper state flux ($j_{43}$) and increasing electron temperature ($T_e$), and the plot is closer to a linear form.

\begin{figure}[htpb]
\includegraphics[scale=0.75]{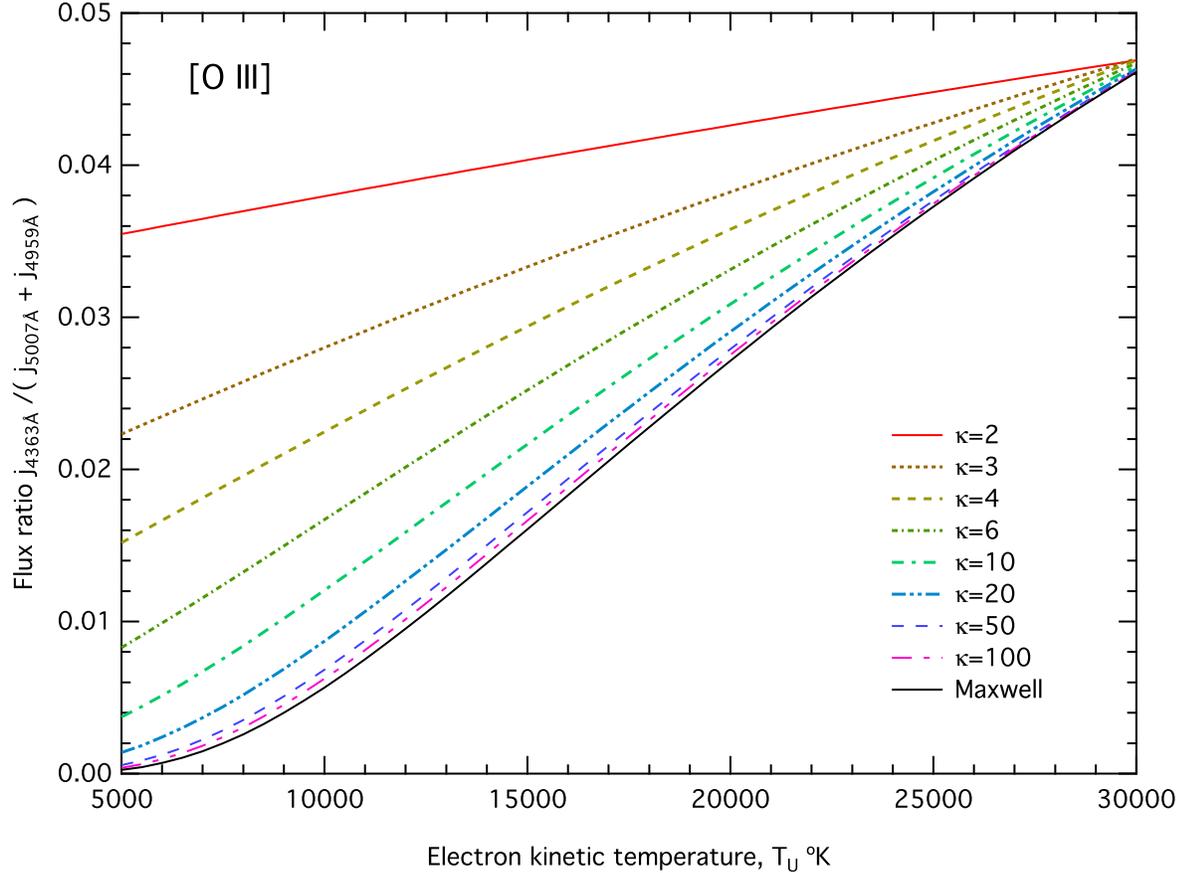}
\caption{Temperature-sensitive line flux ratio curve for the O~{\sc iii} forbidden lines using collision strength data from  \citetalias{Palay12}.}\label{f_9}
\end{figure}

Figure \ref{f_9} shows the line flux ratio plotted against electron temperature for the forbidden transitions of [O {\sc iii}], for several values of $\kappa$.

\begin{figure}[htpb]
\includegraphics[scale=0.75]{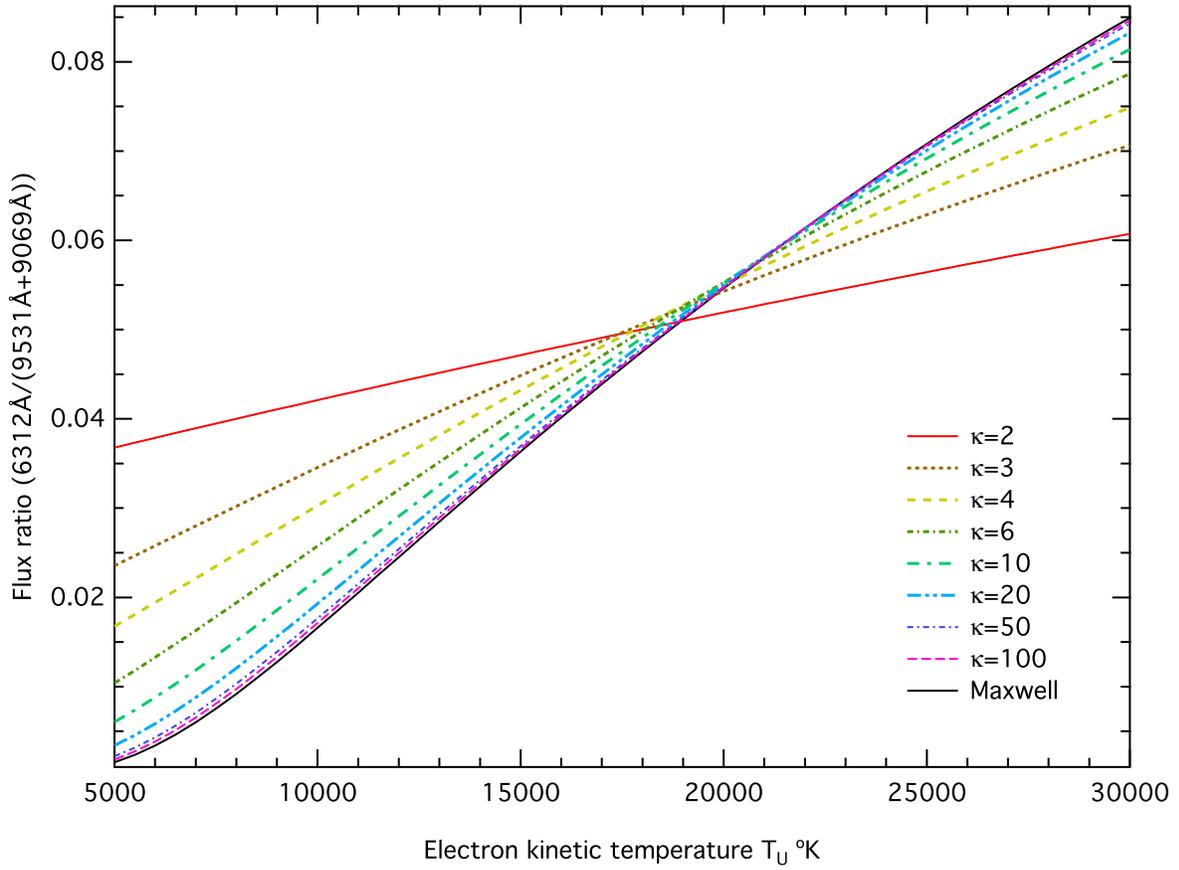}
\caption{Temperature-sensitive line flux ratio curve for the S~{\sc iii} forbidden lines using collision strength data from  \citet{Hudson12}.}\label{f_10}
\end{figure}

\FloatBarrier 
Figure \ref{f_10} is the same as Figure \ref{f_9}, but for the [S~{\sc iii}] transitions.  It differs noticeably from the [O~{\sc iii}] case, owing to the lower excitation energy of the upper state of the 6312\AA\ auroral line. The implication is that in extremely low metallicity, high electron temperature plasmas, above $\sim$20,000K, the effect of the $\kappa$ distribution is to increase the kinetic temperature above the value suggested by assuming a M-B distribution, rather than the reverse which applies to [O~{\sc iii}] transitions for similar metallicity and temperature environments. 

\FloatBarrier 
Although it occurs at different temperatures for different atomic species, this crossover point in the line ratio flux plots appears to be a universal phenomenon, a point on the electron temperature scale where the collisional excitation generates a line flux ratio which is the same for any value of the parameter $\kappa$, including the M-B distribution.\footnote{ A suitable term for the crossover is the `isodierethitic point', from the (ancient) Greek \textgreek{'isos}(equal) and \textgreek{dier'ejisis} (excitation) \citep[reference:][]{Liddell40}}

\subsection{Comparison of [S~{\sc iii}] and [O~{\sc iii}] electron temperatures as a function of $\kappa$}

\begin{figure}[htpb]
\includegraphics[scale=0.75]{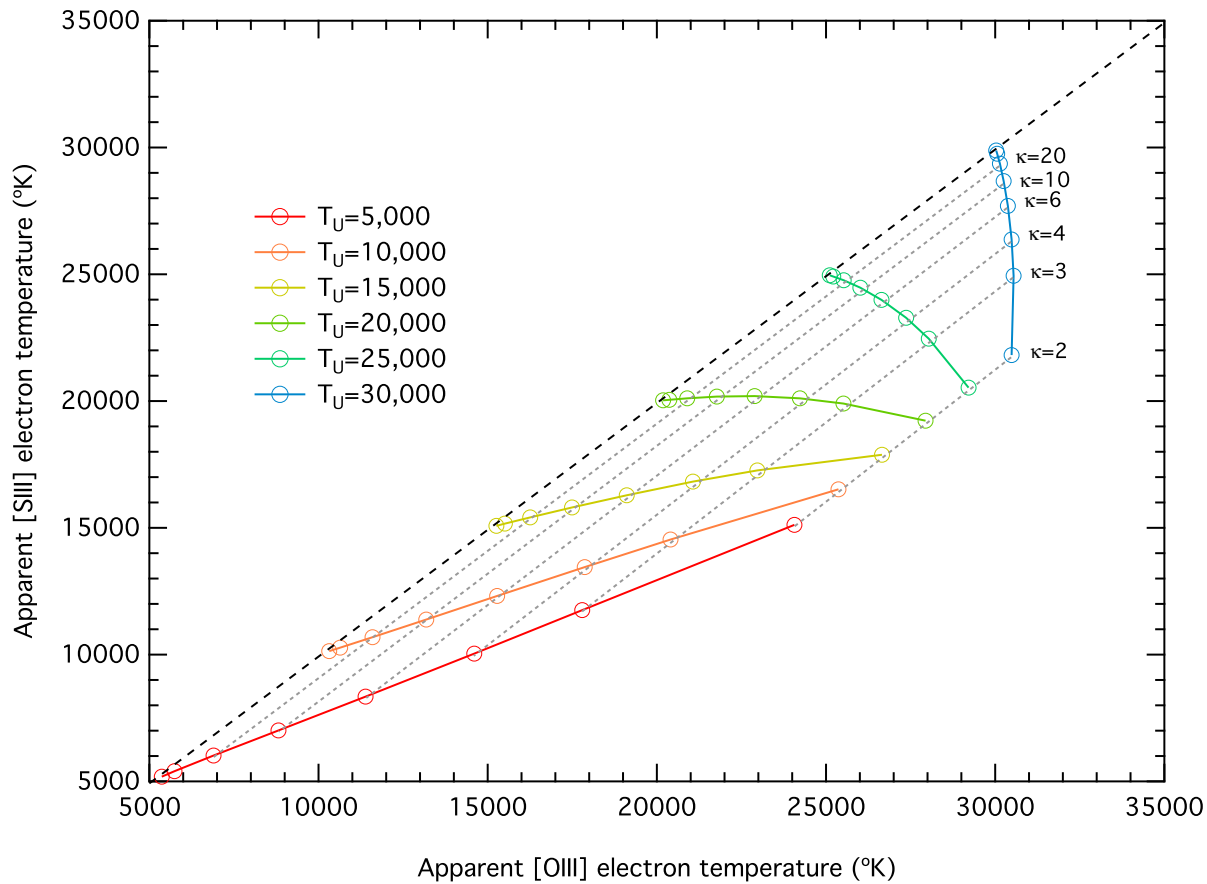}
\caption{Locus of ``apparent'' electron temperatures measured using [S~{\sc iii}] and [O~{\sc iii}] flux ratios, as functions of $\kappa$ and the kinetic (internal energy) temperature, $T_U$. With ``equilibrium'' measurements of electron temperatures from the two species, it is possible to estimate both $\kappa$ and the true kinetic temperature.}\label{f_11}
\end{figure}

Figures \ref{f_9} and \ref{f_10} above show that as $\kappa$ varies, electron temperature measurements using the [S~{\sc iii}] flux ratios will differ from equivalent measurements using the [O~{\sc iii}] lines. Figure \ref{f_11} shows how the two measurements relate to each other, and provides a means of estimating the value of $\kappa$ and $T_U$ by comparing the two measured electron temperatures. Values of the [S~{\sc iii}] flux ratios determined assuming M-B equilibrium are plotted against similar values using the [O~{\sc iii}] flux ratios, as a function of $\kappa$ and the kinetic temperature, $T_U$.  This demonstrates how the $\kappa$ distribution can explain discrepancies between CEL  temperatures from different species.  See also Figure 7 from \citet{Binette12}.

\FloatBarrier 

\section{Tools for measuring the equilibrium and true (kinetic) temperatures}
\subsection{Calculating equilibrium temperatures using the latest collision strengths}

There are three methods commonly used to measure electron temperatures from CEL flux ratios.  The first is to use the simple exponential expression  \citep[equation 5.4, \emph{et seq.},][]{Osterbrock06}, or the equivalent, using the flux ratio/temperature graphs, e.g.,  in Figure \ref{f_10} or the inverse graphs given in \citet{Osterbrock06}.  The second, in the case of O~{\sc iii}, is to use the iterative process described by \citet{Izotov06}.  The third is to use the IRAF STSDAS/nebular/temden routine \citep{Shaw95} or PyNeb \citep{Luridiana12}. If we assume the electrons exhibit a M-B energy distribution, the accuracy of  these methods depends (\emph{inter alia}) on the accuracy of the collision strengths used, and all of these methods make use of older values for the effective collision strengths. For example, IRAF/temden by default uses O~{\sc iii} data from \citet{Lennon94} and O~{\sc ii} energy levels dating from 1960. In many cases, more recent and more accurate atomic data are available, and should be used in preference to older data.

\begin{figure}[htpb]
\includegraphics[scale=0.75]{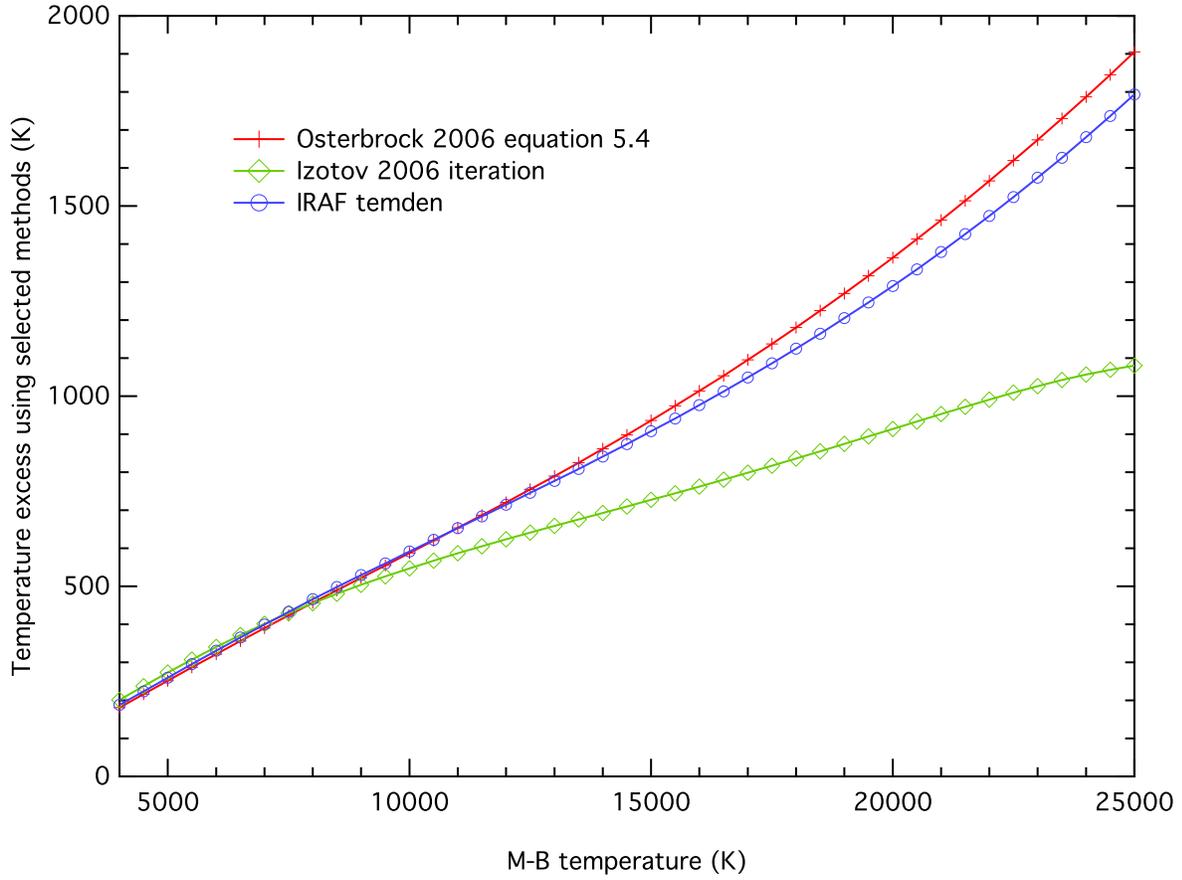}
\caption{Errors arising from using old O~{\sc iii} atomic data in conventional electron temperature estimates (electron density 100 cm$^{-3}$). \emph{c.f.} Figure \ref{f_2}, which uses published effective collision strengths, rather than the numerically integrated collision strengths and resultant flux ratios, used here.}\label{f_12}
\end{figure}

To illustrate the differences that arise from using older data, for O~{\sc iii}, we calculate the flux ratios using the M-B averaged detailed collision strengths from \citetalias{Palay12} for a range of equilibrium temperatures and an electron density of $100\ \text{cm}^{-3}$, and then use these flux ratios as input to the the methods mentioned above.  The results are given in Figure \ref{f_12}. The differences are considerable and point out the errors inherent in using old data.   In this section we present a simple method for calculating equilibrium electron temperatures directly  from observed line flux ratios, using the most recent atomic data.

\FloatBarrier 

Flux ratios of temperature sensitive collisionally excited lines have been used for many years to measure electron temperatures.  Most frequently used is the ratio of the [O~{\sc iii}] nebular and auroral lines, but line flux ratios of several other species have been used.  Table \ref{t_3} lists line flux ratios (for which detailed collision strength data is available) that have been or can be used to estimate electron temperatures.  Some species, for example, S~{\sc ii} and O~{\sc ii}, can be used to estimate both electron densities and temperatures.

The most accurate method for calculating the equilibrium electron temperature from line flux ratios is to compute the flux ratios as a function of temperature by convolving the collision strengths with the M-B distribution, using equation \eqref{e8}.  However, based on these calculations, a much simpler approach is possible, which allows the observer to calculate the M-B temperature directly from the line flux ratio measurements. This involves fitting a simple power law to the computed flux ratio vs equilibrium temperature curves.

An expression involving the flux ratio $\cal{R}$ of the form:
\begin{equation}\label{e32}
T_{\text{MB}} = a\  (-log_{10}({\mathcal{R}}) -b)^{-c}
\end{equation}
gives equilibrium temperatures accurate to within 0.5\% of the computed collision strength values, where the flux ratio
\begin{equation}\label{e33}
\mathcal{R} = \frac{j_{\lambda a}}{j_{\lambda o1}+j_{\lambda o2}}
\end{equation}
used in equation \eqref{e32} is as defined in Tables \ref{t_3} and \ref{t_4}, and the inverse of the ratio used in \citet[eqn. 5.4]{Osterbrock06}. The observer simply uses equation \eqref{e32} with the observed line flux ratio to calculate the electron temperature. The equation coefficients $a, b$, and $c$  are given in Table \ref{t_3}. This method has the advantage that equilibrium electron temperatures can be calculated directly from the observed data, while making use of the latest collision strengths.

The $p^2$ and $p^4$ ions in Table \ref{t_3} are those normally used for electron temperature measurement.  It is quite feasible to use $p^3$ ions, but most of these are also strongly density sensitive, so flux ratios calculated simply from collision strength data for these lines---auroral or UV---will only give useful temperatures at densities $\lesssim$5 cm$^{-3}$. All ratios listed here increase in value as the electron temperature increases (the inverse of the conventional approach).

\begin{table}[htbp]
\caption{Line ratios and M-B temperature fit coefficients from simple collision strength calculations}
\begin{center}
\begin{tabular}{lllll}
\\ \hline \hline
$p^2$, $p^4$ ions & & & & \\
\hline
Species & Line Ratio & a & b & c \\  \hline
$[$O~{\sc i}$]$ & j(5577) / j(6300+6363) & 8488.9 & 0.86645 & 0.9578 \\ 
$[$N ~{\sc ii}$]$ & j(5755) / j(6548+6583) & 11187 & 0.85916 & 1.0259 \\ 
$[$S ~{\sc iii}$]$ & j(6312) / j(9069+9532) & 11237 & 0.67368 & 1.0835 \\ 
$[$O ~{\sc iii}$]$ & j(4363) / j(4959+5007) & 13748 & 0.87704 & 1.0064\\ 
$[$Ne ~{\sc iii}$]$ & j(3342) / j(3869+3969) & 14911 & 1.2619 & 1.0270 \\ 
$[$Ar ~{\sc iii}$]$ & j(5192) / j(7136+7751) & 11899 & 0.96857 & 0.9897 \\ 
 \hline
$p^3$ ions & & & & \\ \hline
Species & Line Ratio & a & b & c \\  \hline
$[$S~{\sc ii}$]$ & j(10287+10321+10336+10371) / j(6716+6731) & 6965.6 & 0.64471 & 0.9960 \\ 
$[$O~{\sc ii}$]$ & j(7319+7320+7330+7331) / j(3726+3729) & 9090.5 & 0.87779 & 1.0161 \\ 
$[$Ar ~{\sc iv}$]$ & j(7171+7238+7263+7332) / j(4711+4740) & 9935.6 & 0.64612 & 1.1243 \\ 
\hline \hline
\end{tabular}
\end{center}
\label{t_3}
\end{table}

However, a more sophisticated approach is possible using the MAPPINGS IV photionization code, which makes use of the latest collision strength and effective collision strength data, and takes into account densities. We discuss this in the following section.

\FloatBarrier 

\subsection{The effect of densities on measured temperatures}

All line ratios are ultimately dependent upon both temperature and density. For temperature sensitive ratios, a number of attempts have been made account for the effect of electron density on the temperatures measured using CEL ratios. For example see \citet[][equations 5.4 through 5.7]{Osterbrock06} and the IRAF/temden routine provides a multi-level approach for the commonly used ions.  Again, these procedures are approximations and/or are based on older atomic data\footnote{See footnote 3: PyNeb is a revised and extended Python-based version of the IRAF  nebular/temden routines, developed by \cite{Luridiana12}}.  Here we have used the newly revised MAPPINGS IV photoionization code to explore how electron density affects computed temperatures.  MAPPINGS IV takes into account the multi-level nature or the atomic species involved in generating the emission lines whose ratios are used to compute electron temperatures. The code uses the latest detailed collision strengths (see Table \ref{t_2}) or the latest available atomic data for effective collision strengths where detailed collision strengths are not available. It also uses a consistent set of transition probabilities \citep{Dopita13}.

\begin{figure}[htpb]
\includegraphics[scale=0.75]{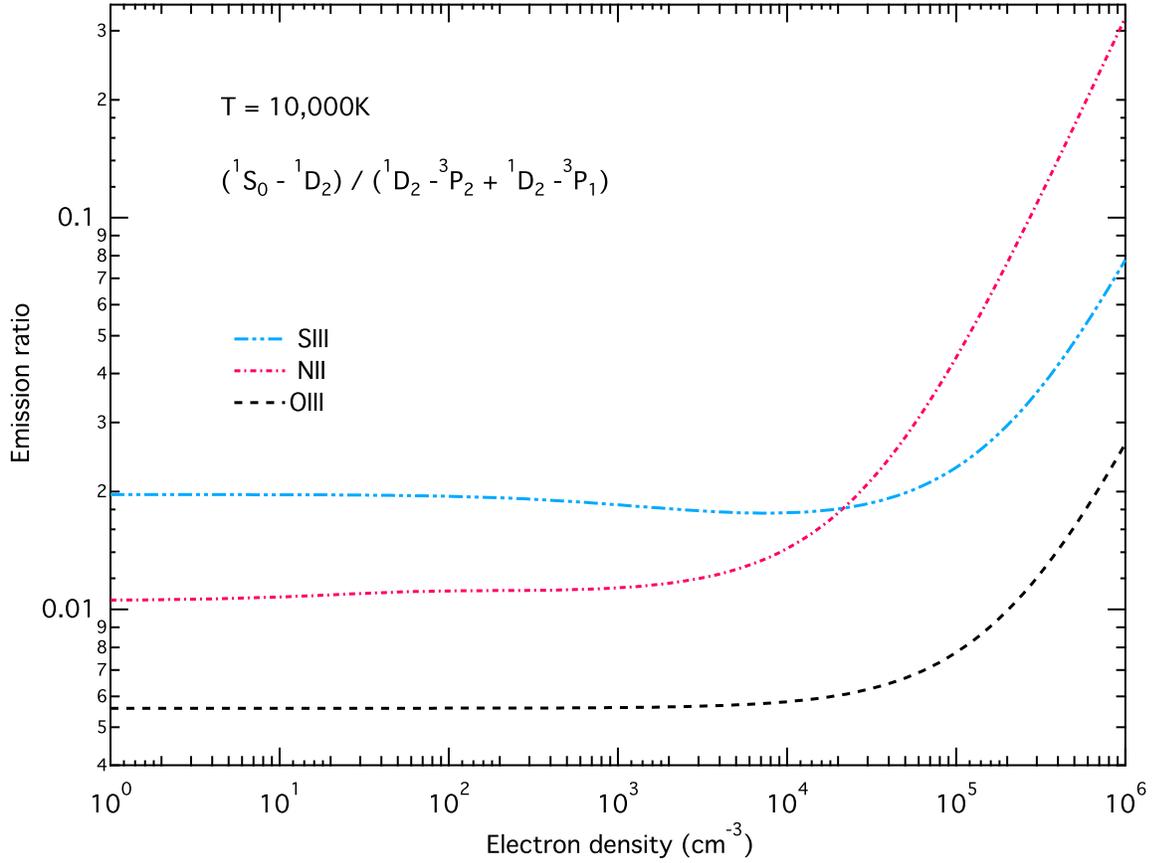}
\caption{Effect of density on the flux emission ratios for the p$^2$ ions, S~{\sc iii}, N~{\sc ii}, and O~{\sc iii}, plotted on a log-log scale. The flux ratio rises at high density due to collisional de-excitation of the $^1$D$_2 - ^3$P transitions. Note the non-constant behaviour of the [N~{\sc ii}] and [S~{\sc iii}] line ratios below $n_e \sim 10^4$. This is due to collisional re-adjustment of the $^3$P levels before they are populated according to their statistical weights.}\label{f13}
\end{figure}

\begin{figure}[htpb]
\includegraphics[scale=0.75]{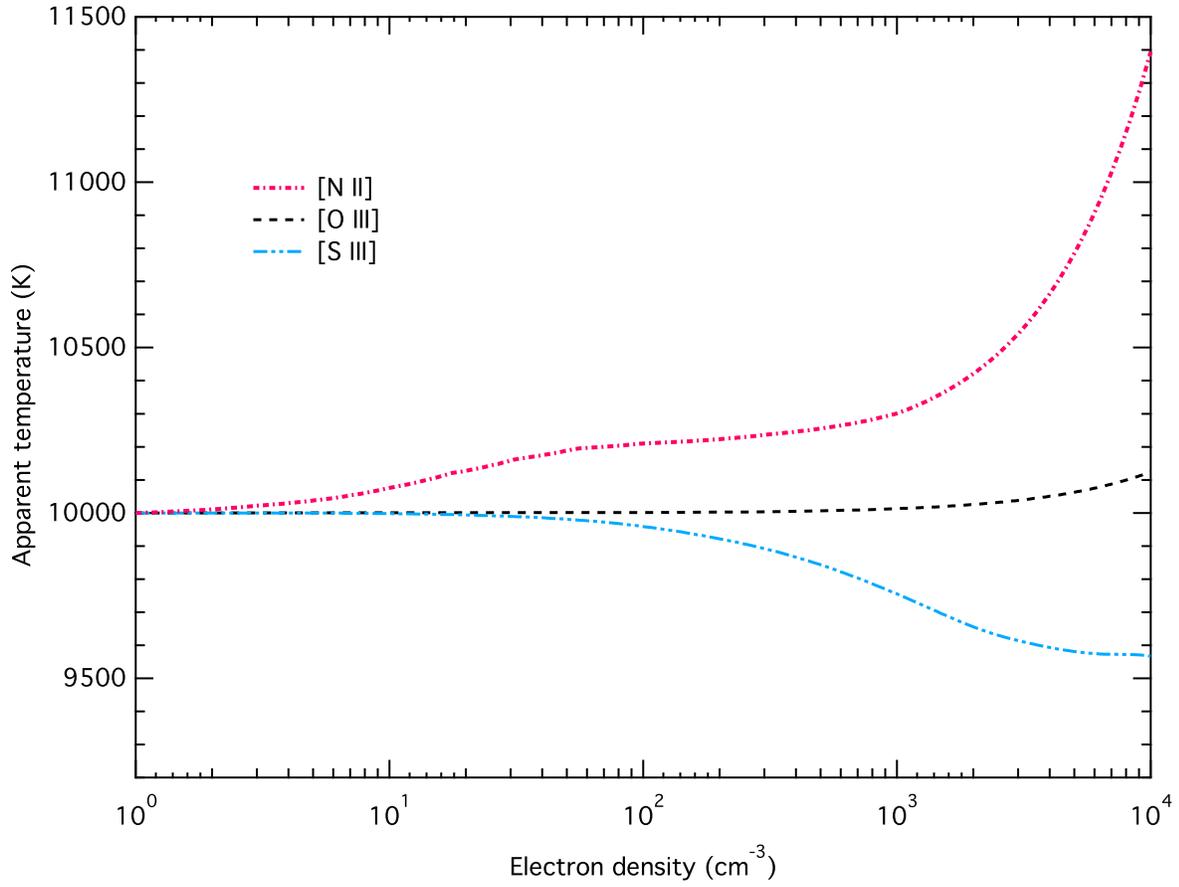}
\caption{Errors arising in the M-B temperature calculations where no correction is made for density, for O~{\sc iii},   S~{\sc iii}, and  N~{\sc ii}, plotted on a linear-log scale. The non-constant behaviour of the [N~{\sc ii}] and [S~{\sc iii}] line ratios is more clearly shown in this graph.}\label{f14}
\end{figure}

Figure \ref{f13} shows the effect of density on the ratio of auroral to optical line fluxes for [S~{\sc iii}], [N~{\sc ii}], and [O~{\sc iii}], calculated using MAPPINGS IV, for a M-B temperature of 10,000K.  Figure \ref{f14} shows what temperature these ratios would imply without any density correction.  It is apparent that, for most ions,  without correction, substantial errors will be made in the in the estimated M-B temperature,  even at moderate densities.

\FloatBarrier 

In a more comprehensive approach to determining the M-B temperatures from ion flux ratios in the presence of changing densities, we  have computed the temperature behavior for several important and widely used line ratios, using MAPPINGS IV, at a range of densities, from 1 to 10$^4$ cm$^{-3}$, and have derived simple linear fits as per equation \eqref{e32}.  The line ratios and the results of these fits are presented in Table \ref{t_4}. Note that most of these ratios use the brightest and most spectroscopically convenient lines likely to be observed in nebular spectra.  In general, we use simpler ratios than those in Table \ref{t_3}, to make use of bright nebular lines and those least sensitive to density effects.  However, the full ratio for [O~{\sc iii}] is also presented for comparison with Table \ref{t_3}. 

\begin{table}[htbp]
\caption{Line ratios and M-B temperature fit coefficients from MAPPINGS IV}
\begin{center}
\begin{tabular}{lllll}
\\ \hline \hline
$p^2$, $p^4$ species &  & \multicolumn{1}{l}{} & \multicolumn{1}{l}{} & \multicolumn{1}{l}{} \\ 
Species & Line Ratio & a & b & c \\  \hline
$[$O~{\sc i}$]$ & j(5577)/j(6300) & 10512 & 0.87725 & 0.92405 \\ 
$[$N ~{\sc ii}$]$ & j(5755)/j(6583) & 10873 & 0.76348 & 1.01350 \\ 
$[$S ~{\sc iii}$]$ & j(6312)/j(9069) & 10719 & 0.09519 & 1.03510 \\ 
$[$S ~{\sc iii}$]$ & j(6312)/j(9069+9532) & 10719 & 0.64080 & 1.03510 \\ 
$[$Ar ~{\sc iii}$]$ & j(5192)/j(7136) & 11887 & 0.98752 & 0.99124 \\ 
$[$O ~{\sc iii}$]$ & j(4363)/j(5007) & 13229 & 0.79432 & 0.98196 \\ 
$[$O ~{\sc iii}$]$ & j(4363)/j(4959+5007) & 13229 & 0.92350 & 0.98196 \\ 
$[$Ne ~{\sc iii}$]$ & j(3342)/j(3869) & 18419 & 1.01660 & 0.99815 \\ 
$[$Ar ~{\sc v}$]$ & j(4626)/j(7006) & 13131 & 0.67472 & 0.98282 \\ 
$[$Ne ~{\sc v}$]$ & j(2975)/j(3426) & 22471 & 1.00700 & 1.08260 \\  \hline
$p^3$ species &  & \multicolumn{1}{l}{} & \multicolumn{1}{l}{} & \multicolumn{1}{l}{} \\  \hline
Species & Line Ratio & a & b & c \\  \hline
$[$S~{\sc ii}$]$ & j(4068+4076)/j(6731) & 5483.8 & 0.25461 & 0.88515 \\ 
$[$S~{\sc ii}$]$ & j(4068+4076)/j(6717+6731) & 5483.8 & 0.65255 & 0.88515 \\ 
$[$O~{\sc ii}$]$ & j(7320+7330)/j(3726+3729) & 7935.2 & 0.98516 & 0.94679 \\ 
$[$Ar ~{\sc iv}$]$ & j(7171)/j(4740) & 12665 & 1.09820 & 1.18100 \\ 
$[$Cl ~{\sc iii}$]$ & j(3342+3358)/j(5517+5538) & 6637.9 & 0.41953 & 0.91886 \\ 
 \hline \hline
\end{tabular}
\end{center}
\label{t_4}
\end{table}

The fit coefficients for [O~{\sc iii}] differ somewhat from those in Table \ref{t_3}, and show the effects of fully modelling excitation balances using multi-level atoms, rather than the simpler approach taken for Table \ref{t_3}. They should be used in preference to  Table \ref{t_3}.

\FloatBarrier 

For all ions with the exception of N~{\sc ii} and S~{\sc iii}, the density effect can be accommodated by the inclusion of a term which quantifies the collisional de-excitation of the middle level.  This takes the form used by \citet{Osterbrock06}:

\begin{equation}\label{e34}
\mathcal{R} = \frac{\mathcal{R}_{obs}}{1+d\ (N_e/{T}^{1/2})} \ ,
\end{equation}
where $d$ is a constant related to the critical density for the transition,  $n_{\rm crit} = (T^{1/2}/d)$. $\mathcal{R}$ is the ``corrected'' value of the observed density $\mathcal{R}_{obs}$, such that the calculated temperature is the true M-B temperature.  Because the density effects are complex, it is necessary in some cases to use two different values of the parameters for different density ranges.  Table \ref{t_5} shows the values of $d$ for different density ranges, and for different species.

For N~{\sc ii} and S~{\sc iii} a more complex form must be chosen, since the collisional re-adjustment of the $^3$P levels with increasing density causes the peculiar behaviour seen in Figure \ref{f13}. For  N~{\sc ii} an excellent fit can be obtained with two separate values of $a$, $b$ and $d$, applicable over different density ranges, as indicated in the footnote to Table \ref{t_5}. 

Combining equations \eqref{e32} and \eqref{e34}, one can use the $a, b, c,$ and $d$ parameters from Tables \ref{t_4} and \ref{t_5} to fit both densities and temperatures with a single equation,
\begin{equation}\label{e35}
T_{\text{MB}} = a\ \left[-log_{10}\left(  {\frac{\mathcal{R}_{obs}}{1+d\ (N_e/{T}^{1/2})} }\right) -b\ \right] ^{-c }\ .
\end{equation}

\begin{landscape}
\begin{table}[htbp]
  \caption{Density parameter $d$ used in equations \eqref{e34} and \eqref{e35}.}
    \begin{tabular}{l|cccccccccccc}
    \\ \hline \hline
          & O~{\sc iii}$^{a}$ & S~{\sc iii} & N~{\sc ii}$^{b}$ & O~{\sc i} & S~{\sc ii} & O~{\sc ii} & Ne~{\sc iii} & Ar~{\sc iv}$^{c}$ \\
    Log density (cm$^{-3}$) &       &       &       &       &       &       &       &  \\
    \hline
    0.0 $<$ $log_{10} (N_e)$ $<$ 2.0 & 3.8895E-04 & -6.50E-03 & 5.80E-02 & 2.20E-04 & 3.90E-02 & 1.05E-01 & 0.00E+00 & 2.90E-02 \\
    2.0 $<$ $log_{10} (N_e)$ $<$ 3.0 & 3.8895E-04 & -6.50E-03 & 3.60E-03 & 2.20E-04 & 3.90E-02 & 8.90E-02 & 0.00E+00 & 2.90E-02 \\
    3.0 $<$ $log_{10} (N_e)$ $<$ 3.5 & 3.8895E-04 &   n/a$^{d}$    & 3.60E-03 & 2.20E-04 & 3.90E-02 & 8.90E-02 & 0.00E+00 & 2.90E-02 \\
    3.5 $<$ $log_{10} (N_e)$ $<$ 4.0 & 3.8895E-04 &   n/a$^{d}$    & 3.60E-03 & 2.20E-04 &   n/a$^{d}$    &   n/a$^{d}$    & 0.00E+00 & 2.90E-02 \\
    \hline \hline
    \end{tabular}
   \flushleft
   {a. The fit with $d$= 3.8895 is applicable up to $log_{10} (N_e) < 5.3$ with an error of less than $10^{-4}$.  \newline
   b. Better fits to the density behaviour can be obtained for N~{\sc ii} with the following parameters: \newline
  $0.0 < log_{10} (N_e) <1.5$: $a$=10850, $d$=1.45E-01; $1.5 < log_{10} (N_e) < 5.2$: $a$=10820, $b$=0.8762, $d$=3.0E-03.\newline
   c. Better fits to the density behaviour can be obtained for Ar~{\sc iv} with the following $a$ parameters: \newline
   $log_{10} (N_e) <3.0$: $a$=12665,  $4 > log_{10} (N_e) > 3.0$: $a$=14551 (error $ < 2\times10^{-3}$).
   \newline
   d. ``n/a'' means that the fits are not relaible for these species at these densities, so the parameter \emph{d} is not available, and equations \eqref{e34} and \eqref{e35} are not applicable in these situations.
   }
  \label{t_5}
\end{table}
\end{landscape}

\subsection{Calculating $\kappa$ dependence}
In the above approach, we assume that the electron energies are in thermal equilibrium. No insight is given into the effects of non-equilibrium electron energies. To take the effects of a $\kappa$ distribution into account, we can use Figures \ref{f_9} and \ref{f_10} to measure graphically  the true kinetic temperature from the [O~{\sc iii}] and [S~{\sc iii}] CEL flux ratios for a range of values of the parameter $\kappa$. Similar graphs may be derived for other CEL species.  However, an easier method is to derive a simple linear equation from the graph, that expresses the kinetic temperature in terms of the temperature measured using conventional M-B methods, such as the formula in equation \eqref{e32}.  This is based on the near-linearity of the curves in Figures \ref{f_9} and \ref{f_10}  for temperatures between 4,000K and 25,000K. 

\begin{figure}[htpb]
\includegraphics[scale=0.75]{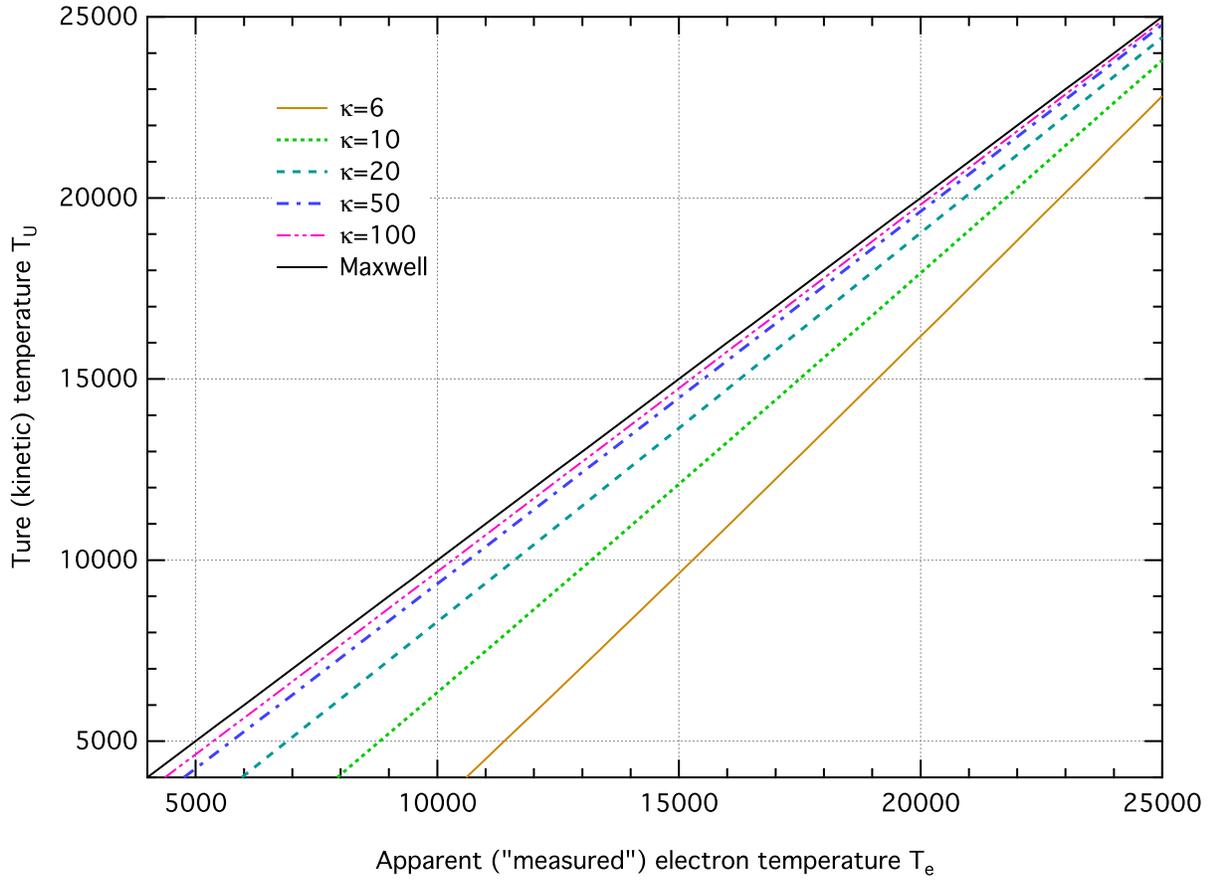}
\caption{True kinetic temperature vs.\ electron temperature for [O {\sc iii}] estimated  using conventional methods. It shows the effect of the parameter $\kappa$ in reducing the true kinetic temperature.}\label{f_13}
\end{figure}

For the range of temperatures (4,000 $< T_U <$ 25,000K) encountered in \HII regions and many PNe, the relationship between the apparent (M-B) electron temperature $T_e$ and the kinetic temperature $T_U$ can be expressed to very good accuracy as a linear equation with parameters that are quadratic functions of 1/$\kappa$, for all values of $\kappa$, as follows:
\begin{equation}\label{e36}
T_U = a(\kappa)\ T_e + b(\kappa)
\end{equation}
where
\begin{equation}\label{e37}
a = \left(a_1+\frac{a_2}{\kappa} + \frac{a_3}{\kappa ^2}\right)
\end{equation}
and
\begin{equation}\label{e38}
b =- \left(b_1+\frac{b_2}{\kappa} + \frac{b_3}{\kappa ^2}\right)
\end{equation}
and where $T_e$ is  derived from conventional equilibrium methods such as equation \eqref{e32}. The  equation coefficients can be derived for any CEL species for which non-averaged collision strengths are available. For the [O {\sc iii}] CELs, this equation is illustrated graphically for a range of values of $\kappa$ in Figure \ref{f_13}. The parameters $a_1$, $a_2$, $a_3$, $b_1$, $b_2$,  and $b_3$ are given in Table \ref{t_6}, for several nebular atomic species.

Using the revised [O~{\sc iii}] atomic data and a $\kappa$ of 10, we see that an apparent [O~{\sc iii}] electron temperature of 15,000K derived via the IRAF/temden routine (with old atomic data) corresponds to a kinetic (internal energy) temperature of $\sim$11,000K.

\FloatBarrier 

\begin{table}[H]
\caption{Linear coefficients for 4,000 $< T_U <$ 25,000K}
\begin{center}
\begin{tabular}{|lll|lll|}
\hline
O {\sc iii} &  &  &  &  &  \\
$a_1$ & $a_2$ & $a_3$ & $b_1$ & $b_2$ & $b_3$ \\ 
1.00036  &  1.27142  &  3.55371  &  21.1751  &  42693.5  &  103086 \\ \hline
S {\sc iii} &  &  &  &  &  \\
$a_1$ & $a_2$ & $a_3$ & $b_1$ & $b_2$ & $b_3$ \\ 
1.00075  &  1.09519  &  3.21668  &  13.3016  &  24396.2  &  57160.4 \\ \hline
N {\sc ii} &  &  &  &  &  \\
$a_1$ & $a_2$ & $a_3$ & $b_1$ & $b_2$ & $b_3$ \\ 
1.0008  &  1.26281  &  3.06569  &  19.432  &  31701.9  &  70903.4 \\ \hline
O {\sc i} * &  &  &  &  &  \\
$a_1$ & $a_2$ & $a_3$ & $b_1$ & $b_2$ & $b_3$ \\ 
0.995398 & 1.02877 & 2.80919 & -45.0544 & 38145.8 & 92736.3 \\ \hline 
Ne {\sc iii} &  &  &  &  &  \\
$a_1$ & $a_2$ & $a_3$ & $b_1$ & $b_2$ & $b_3$ \\ 
1.00128  &  1.18331  &  2.01139  &  42.7545  &  50728.7  &  100311 \\ \hline
Ar {\sc iii} &  &  &  &  &  \\
$a_1$ & $a_2$ & $a_3$ & $b_1$ & $b_2$ & $b_3$ \\ 
1.00087  &  1.4561  &  3.62825  &  17.4001  &  32190  &  72395 \\ \hline
Ar {\sc v} &  &  &  &  &  \\
$a_1$ & $a_2$ & $a_3$ & $b_1$ & $b_2$ & $b_3$ \\ 
1.00083  &  1.10823  &  3.15458  &  19.6749  &  31483.4  &  72117.8 \\ \hline
O~{\sc ii} &  &  &  &  &  \\
$a_1$ & $a_2$ & $a_3$ & $b_1$ & $b_2$ & $b_3$ \\ 
 0.999828  &  1.36062  &  2.99473  &  19.5715  &  47145  &  107673 \\ \hline 
S~{\sc ii} &  &  &  &  &  \\
$a_1$ & $a_2$ & $a_3$ & $b_1$ & $b_2$ & $b_3$ \\ 
 0.998299  &  1.62932  &  3.9533  &  -17.6454  &  31513  &  73137.3 \\ \hline 
Ar~{\sc iv} &  &  &  &  &  \\
$a_1$ & $a_2$ & $a_3$ & $b_1$ & $b_2$ & $b_3$ \\ 
 0.99791  &  1.19881  &  3.82751  &  -13.2742  &  36490.7  &  98591.4\\ \hline
\end{tabular}
\end{center}
\label{t_6}
\end{table}
* For O~{\sc i}, the linear range is between 1,000K and 20,000K.
\FloatBarrier 

\section{Strong line techniques}
Numerous methods have been developed using ratios of the strong lines in nebular spectra, which are important in the absence of direct electron temperature diagnostic lines.  See, for example, \citet{Kewley08} and \citet{ Kewley02}.  These methods make use of the line fluxes from a range of different atomic species, usually selected because they are readily measurable with low noise in most nebular spectra. However the impact of a $\kappa$ distribution on these methods is not simple, as each species is affected to a different extent by the distribution. It is necessary to calculate and model each strong line index separately as a function of temperature. Initial investigations suggest that several of the methods will not be strongly affected by $\kappa$ distributions, and in particular, measurements comparing [S~{\sc ii}]  6716\AA, 6731\AA\ and [N~{\sc ii}] 6548\AA, 6583\AA\ are not significantly affected, as the fluxes of both species are changed to a similar extent by a $\kappa$ distribution.

As a simple illustration, we can consider the strong line ratio ``$R_{23}$''.  This flux ratio is given by:
\begin{equation}\label{e39}
R_{23} = ([\text{O~} \textsc{ii}] \lambda 3726 + [\text{O~} \textsc{ii}] \lambda 3729+[\text{O~} \textsc{iii}] \lambda 4959 + [\text{O~} \textsc{iii}] \lambda 5007)/H\beta
\end{equation}

The excitation temperatures for the [O~{\sc ii}] and [O~{\sc iii}] lines are $\sim$38,600K and $\sim$29,000K respectively.  For a $\kappa$ value of 10, and a kinetic temperature of 10,000K, from Figure \ref{f_7}, it is apparent that the [O {\sc ii}] lines are enhanced by $\sim$20\%; the [O~{\sc iii}] lines are not significantly affected; and, from the discussion earlier, H$\beta$ is enhanced by $\sim$4\%.  Thus the overall $R_{23}$ ratio is slightly enhanced.   A detailed analysis of strong line methods is best tackled using photoionization models that take into account$\kappa$ effects. While a detailed analysis of the impacts of $\kappa$ distributions and new atomic data on strong line methods is beyond the scope of the present paper, it is explored in our next paper in this series \citep{Dopita13}, which develops new strong line diagnostics that give significantly more consistent results when compared to direct $T_e$ methods. The subject will be addressed further in subsequent papers.  Nonetheless, it is apparent from this simple example that the effect of changes in the collision strengths and $\kappa$ on derived strong line abundances is relatively small, but not insignificant.

\section{Estimating $\kappa$}

The kappa distribution uses a single parameter to describe the deviation from thermal equilibrium in electron energies.  In any one temperature or abundance measurement, there is no unique way to estimate the value of $\kappa$, although a value of $\sim$10 appears consistent with many of the observed spectra \citepalias{Nicholls12}.  When more than one measurement is available---for example, electron temperatures obtained using different CEL species, or CEL and ORL-derived abundances---the value of kappa can be estimated by the requirement that the discrepancies be minimized.  When several different methods are available such as in bright nebulae, it is possible to iterate to an optimum value of kappa and estimate errors and variance. Figure \ref{f_11} shows that measuring apparent (M-B) electron temperatures for [S~{\sc iii}] and [O~{\sc iii}] allows one to estimate both $\kappa$ and the kinetic (internal energy) temperature.

Needless to say, in real nebulae there are likely to be kappa distributions spanning a range of values of $\kappa$, so specifying a single value is not always meaningful, but the concept can help to avoid the large discrepancies that arise using equilibrium methods, and can augment values obtained using other contributing factors such as temperature and abundance inhomogeneities.

There will seldom be a single answer for temperature, abundances and $\kappa$ for any real nebula, and using photoionization models to explore the complex physics is critically important.  For this reason, we have revised the MAPPINGS III photoionization code \citep{Allen08} to version IV, to incorporate both non-equilibrium $\kappa$ effects and the most accurate available collision strengths and other atomic data. This work is the subject of our next paper  \citep{Dopita13}, where we use it to investigate the effect of $\kappa$ distributions on temperatures and abundances estimated using the strong line methods, to develop a revised set of strong line diagnostics. The code development has been undertaken independently of the work on MAPPINGS Ie  \citep{Binette12}, with which it shares a common origin but which has had a separate development.

\section{Conclusions}

In this paper we have explored further the ideas put forward in \citetalias{Nicholls12}, where the non-equilibrium $\kappa$ electron energy distribution widely encountered in solar system plasmas was found to explain the long standing abundance discrepancy problem that arises when temperatures and abundances are measured using spectra from different atomic species.  We have discussed the factors involved in obtaining accurate CEL temperatures from theoretical collision strengths. We have also shown that significant errors in electron temperatures can arise unless one has access to the best possible collision strength data. We have examined the effects of the $\kappa$ distribution on recombination processes, in particular how the $\kappa$ distribution is able to resolve the long standing discrepancy between ORL and CEL abundances.  We show that a typical $\kappa$ distribution leads to a small enhancement of hydrogen recombination lines. We have examined in detail the effects of $\kappa$ and newly available collision strength data affects the measurement of electron temperatures using collisionally excited lines.  We compare these effects on the forbidden lines of S~{\sc iii} and O~{\sc iii}. In the main thrust of the paper, we present simple techniques for calculating equilibrium electron temperatures from line flux ratios using the most up to date atomic data, and using these equilibrium temperatures to derive the actual kinetic (internal energy) temperatures resulting from non-equilibrium electron energy distributions. We outline future work on adapting photoionization modelling programs and strong line methods to take into account the effects of the $\kappa$ distribution.

\appendix
\section {Supplementary Data Tables}

\subsection{Temperature-sensitive line ratio data}
The tables in this appendix give the wavelengths, transition probabilities and line ratio multipliers (equations \ref{e29} and \ref{e31}), for transitions of the $p^2$, $p^4$ and $p^3$ lines of nebular interest.  For the meanings of the wavelength and transition probability symbols, see sections 5.2 and 5.3.

\begin{table}[htbp]
\caption{Line wavelengths (\AA) (in air), line strengths and line ratio multipliers for the $p^2$ and $p^4$ ions.  The final line in this table shows the $f_2$(A,$\lambda$) line ratio multiplier for the UV-to-optical line ratios.  They are related to the auroral-to-optical line ratios via the wavelength weighted branching ratios.}
\begin{tabular}{llllllll}
\\ \hline \hline
Species & \multicolumn{1}{c}{O~{\sc i}} & \multicolumn{1}{c}{N~{\sc ii}} & \multicolumn{1}{c}{O~{\sc iii}} & \multicolumn{1}{c}{S~{\sc iii}} & \multicolumn{1}{c}{Ne~{\sc iii}} & \multicolumn{1}{c}{Ar~{\sc iii}} & \multicolumn{1}{c}{Ar~{\sc v}} \\ 
\hline
$\lambda_{a}$(\AA) & 5577.3 & 5754.6 & 4363.2 & 6312.1 & 3342.2 & 5191.8 & 4625.4 \\ 
$\lambda_{o2}$(\AA) & 6363.8 & 6548.0 & 4958.9 & 9068.6 & 3968.5 & 7751.1 & 6435.1 \\ 
$\lambda_{o3}$(\AA) & 6300.3 & 6583.4 & 5006.8 & 9530.6 & 3868.8 & 7135.8 & 7005.8 \\ 
$\lambda_{u2}$(\AA) & 2972.3 & 3062.8 & 2321.0 & 3721.6 & 1814.6 & 3109.2 & 2691.1 \\ 
$\lambda_{u3}$(\AA) & 2958.4 & 3070.6 & 2331.4 & 3797.2 & 1823.7 & 3005.2 & 2786.0 \\ 
$A_{a}$ & 1.26E+00 & 1.14E+00 & 1.71E+00 & 2.08E+00 & 2.65E+00 & 3.10E+00 & 3.80E+00 \\ 
$A_{o2}$ & 1.82E-03 & 9.84E-04 & 6.21E-03 & 1.85E-02 & 5.40E-02 & 8.31E-02 & 2.26E-01 \\ 
$A_{o3}$ & 5.65E-03 & 2.91E-03 & 1.81E-02 & 4.80E-02 & 1.74E-01 & 3.35E-02 & 5.20E-01 \\ 
$A_{u2}$ & 7.54E-02 & 3.18E-02 & 2.15E-01 & 6.61E-01 & 2.06E+00 & 4.02E+00 & 6.80E+00 \\ 
$A_{u3}$ & 2.42E-04 & 1.55E-04 & 6.34E-04 & 8.82E-03 & 4.00E-03 & 4.30E-02 & 8.10E-02 \\ 
$f_1$(A,$\lambda$) & 1.06824 & 1.11132 & 1.01650 & 1.12615 & 0.65458 & 0.63050 & 0.52476 \\ \hline
$f_1$(A,$\lambda$)uv & 0.12034 & 0.05853 & 0.24096 & 0.61492 & 0.93727 & 1.38038 & 1.63260 \\ \hline
\hline
\end{tabular}
\label{t_7}
\end{table}

\begin{table}[htbp]
\caption{Line wavelengths (\AA) (in air), line strengths and line ratio multipliers for the $p^3$ ions.  Alternative flux ratios can be used for the $p^3$ ions, using the UV lines in place of the ``auroral''.  This is done, for example, in the IRAF/temden routine for S~{\sc ii}, Ne~{\sc iv} and Ar~{\sc iv}. The final line in this table shows the $f_2$(A,$\lambda$) line ratio multiplier for the UV-to-optical line ratios.  They are related to the auroral-to-optical line ratios via the wavelength weighted branching ratios.}
\begin{center}
\begin{tabular}{llllll}
\hline
\hline
Species & \multicolumn{1}{c}{N~{\sc i}}& \multicolumn{1}{c}{S~{\sc ii}} & \multicolumn{1}{c}{O~{\sc ii}} & \multicolumn{1}{c}{Ne~{\sc iv}} & \multicolumn{1}{c}{Ar~{\sc iv}} \\ 
\hline
$\lambda_{a1}$ & 10407.6 & 10370.5 & 7330.7 & 4725.6 & 7332.2 \\
$\lambda_{a2}$ & 10407.2 & 10336.4 & 7329.7 & 4724.2 & 7263.3 \\ 
$\lambda_{a3}$ & 10398.1 & 10320.5 & 7320.0 & 4715.7 & 7237.8 \\ 
$\lambda_{a4}$ & 10397.7 & 10286.7 & 7318.9 & 4714.2 & 7170.7 \\ 
$\lambda_{o1}$ & 5200.3 & 6730.8 & 3728.8 & 2424.4 & 4740.1 \\ 
$\lambda_{o2}$ & 5197.9 & 6716.4 & 3726.0 & 2421.8 & 4711.3 \\ 
$\lambda_{u1}$ & 3466.5 & 4076.3 & 2470.3 & 1601.1 & 2868.2 \\ 
$\lambda_{u2}$ & 3466.5 & 4068.6 & 2470.2 & 1600.9 & 2853.7 \\ 
$A_{a1}$ & 5.31E-02 & 6.81E-02 & 5.34E-02 & 3.89E-01 & 1.22E-01 \\ 
$A_{a2}$ & 2.74E-02 & 1.42E-01 & 8.67E-02 & 4.36E-01 & 6.78E-01 \\ 
$A_{a3}$ & 3.45E-02 & 1.57E-01 & 9.91E-02 & 1.10E-01 & 6.70E-01 \\ 
$A_{a4}$ & 6.12E-02 & 1.15E-01 & 5.19E-02 & 3.01E-01 & 9.08E-01 \\ 
$A_{o1}$ & 7.56E-06 & 6.84E-04 & 3.06E-05 & 5.80E-04 & 7.71E-02 \\ 
$A_{o2}$ & 2.03E-05 & 2.02E-04 & 1.78E-04 & 5.47E-03 & 9.60E-03 \\ 
$A_{u1}$ & 6.50E-03 & 7.72E-02 & 5.22E-02 & 5.30E-01 & 9.70E-01 \\ 
$A_{u2}$ & 2.60E-03 & 1.92E-01 & 2.12E-02 & 1.33E+00 & 2.55E+00 \\ 
$f_2$(A,$\lambda$) & 0.47521 & 0.41815 & 0.40631 & 0.20480 & 0.26438 \\ \hline
$f_2$(A,$\lambda$)uv & 0.07365 & 0.59216 & 0.30377 & 0.90891 & 0.98929  \\ \hline
\hline
\end{tabular}
\end{center}
\label{t_8}
\end{table}

\FloatBarrier 

\begin{acknowledgments}
\section{Acknowledgements}
The authors wish to thank Dr P. Barklem for the collision strength data on O~{\sc i}, Dr C. Ramsbottom for the collision strength data for S~{\sc iii}, N {\sc ii} and Ar {\sc iii};  and Dr S. Tayal  for the collision strength data for O {\sc ii} and N~{\sc ii}. Additional collision strength data is available on the APARC website at http://web.am.qub.ac.uk/apa and from The Iron Project database (TIPbase) at http://cdsweb.u-strasbg.fr/tipbase. M. Dopita acknowledges the support of the Australian Research Council (ARC) through Discovery  project DP0984657.  The authors also wish to thank the anonymous referee for detailed comments and helpful suggestions.
\end{acknowledgments}

\bibliographystyle{aa}
\end{document}